\documentclass[12pt, onecolumn]{IEEEtran} 
\IEEEoverridecommandlockouts
\usepackage{cite}
\usepackage{amsmath,amssymb,amsfonts}
\usepackage{algorithmic}
\usepackage{graphicx}
\usepackage{textcomp}
\usepackage{xcolor}
\usepackage{algorithm}
\usepackage{comment}
\usepackage{setspace}
\usepackage{color}
\usepackage{subfigure}
\newtheorem{theorem}{\bf Theorem}%[section]
\newtheorem{lemma}{\it Lemma}
\newcommand{\bb}{\boldsymbol{b}}
\newcommand{\bx}{\boldsymbol{x}}

\newcommand{\bz}{\boldsymbol{z}}

\newcommand{\bF}{\boldsymbol{F}}
\newcommand{\bmu}{\boldsymbol{\mu}}
\newcommand{\bLambda}{\boldsymbol{\Lambda}}
\newcommand{\bPi}{\boldsymbol{\Pi}}
\newcommand{\bV}{\boldsymbol{V}}
\newcommand{\bv}{\boldsymbol{v}}
\newcommand{\cG}{{\cal G}}

\def\BibTeX{{\rm B\kern-.05em{\sc i\kern-.025em b}\kern-.08em
    T\kern-.1667em\lower.7ex\hbox{E}\kern-.125emX}}
\ifCLASSINFOpdf

\else
\fi
\begin{document}
\title{Federated Orchestration for Network Slicing of Bandwidth and Computational Resource}
\author{Yingyu~Li,
        Anqi~Huang,
         Yong~Xiao,~\IEEEmembership{Senior~Member,~IEEE},
        Xiaohu~Ge,~\IEEEmembership{Senior~Member,~IEEE},
        Sumei~Sun,~\IEEEmembership{Fellow,~IEEE},
        and Han-Chieh~Chao,~\IEEEmembership{Senior~Member,~IEEE}
        \vspace{-1cm}

\thanks{Part of the results in this paper will be presented at the IEEE ICC, Dublin, Ireland, June 2020 \cite{Anqi2020ICCNetSlic}.

Y. Li, A. Huang, Y. Xiao and X. Ge are with the School of Electronic Information and Communications, Huazhong University of Science and Technology, Wuhan 430074 China (e-mails: \{liyingyu, huanganqi, yongxiao\}@hust.edu.cn and xhge@mail.hust.edu.cn).
S. Sun is with the Communications and Networks Cluster, Institute for Infocomm Research, Singapore 138632 (e-mail: sunsm@i2r.a-star.edu.sg).
H.-C. Chao is with the Department of Computer Science and Information Engineering, National Ilan University, Yilan 26047, Taiwan, and also with the Department of Electrical Engineering, National Dong Hwa University, Hualien 97401, Taiwan (e-mail: hcchao@gmail.com). (Corresponding author: Xiaohu Ge.)}
}
\maketitle

\begin{abstract}

Network slicing has been considered as one of the key enablers for 5G to support diversified IoT services and application scenarios. This paper studies the distributed network slicing for a massive scale IoT network supported by  5G  with fog computing. Multiple services with various requirements need to be supported by both spectrum resource offered by 5G network and computational resourc of the fog computing network. We propose a novel distributed framework based on a new control plane entity, federated-orchestrator (F-orchestrator), which can coordinate the spectrum and computational resources without requiring any exchange of the local data and resource information from BSs. We propose a distributed resource allocation algorithm based on Alternating Direction Method of Multipliers with Partial Variable Splitting (DistADMM-PVS). We prove DistADMM-PVS minimizes the average service response time of the entire network with guaranteed worst-case performance for all supported types of services when the coordination between the F-orchestrator and BSs is perfectly  synchronized. Motivated by the observation that coordination synchronization may result in high coordination delay that can be intolerable when the network is large in scale, we propose a novel asynchronized ADMM (AsynADMM) algorithm. We prove that AsynADMM can converge to the global optimal solution with improved scalability and negligible coordination delay. We evaluate the performance of our proposed framework using two-month of traffic data collected in a  in-campus smart transportation system supported by a 5G network. Extensive simulation has been conducted for both pedestrian and vehicular-related services during peak and non-peak hours. Our results show that the proposed framework offers significant reduction on service response time for both supported services, especially compared to network slicing with only a single resource.
\end{abstract}


\maketitle

\IEEEpeerreviewmaketitle

\section{Introduction}

The Internet-of-Things (IoT) is a holistic framework that can support wireless connectivity of a large number of smart devices s such as smart sensors, smart meters, wearable devices, radar, and Light Detection and Ranging (LiDar) systems.  It has been considered as one of the key technologies to fulfill 5G's vision of ubiquitous connectivity. IoT has been originally introduced by various Standard Development Organizations (SDOs) and industries (e.g., LoRa, NB-IoT, SigFox, etc.) as a  low-cost and low-power networking technology with low or limited requirement \cite{3GPPIoT}. It has  been quickly applied and extended into a much broader range of vertical industries with more stringent requirements including Industrial IoT (IIoT) \cite{Taleb2019NetworkSlicing}, Internet of Vehicles (IoV), smart grid networks, etc. According to  ITU-T, {\em IoT is a global infrastructure that could enable diverse advanced services by interconnecting (physical and virtual) ``things" based on existing and evolving inter-operatable  information and communication technologies} \cite{ITU-T2016IoT}.

According to the recently published Network Index report, the total number of connected IoT devices is expected to reach 41.6 billion by 2025. Since there are no resources including both spectrum and computational resources being exclusively assigned for IoT services, how to support a large number of IoT devices with various requirements is one of the main challenges for the next generation networking systems, especially the 5G systems.

%It is commonly believed that 5G will be much more than a simple upgrade of physical performance metrics such as throughput and capacity
5G represents a fundamental transformation from the traditional data-oriented architecture towards a more flexible and service-oriented architecture\cite{NGMN5GWhitePaper} \cite{ge20165g}. The {\em service-based architecture} (SBA) has been introduced by 3GPP  as the key enabler for supporting a plethora of different services with diverse requirements on a common set of physical network resources. The core idea is to use software-defined networking (SDN) and network functions virtualization (NFV) to virtualize the network elements into network functions (NF)s \cite{ETSINFV2017}, each of which consists of a functional building block utilizing various resources offered by the network. Each type of services can then be instantiated by a series of NF sets, called {\em network slice} \cite{NGMN2016NetworkSlicing}. Network slicing has been considered as the foundation of 5G SBA to match with diversified service requirements and application scenarios \cite{Wang2018NetworkSlice}\cite{Sciancalepore2019NetworkSliceBroker}\cite{Samdanis2016Broker}.

To support emerging computationally intensive applications, create new business opportunities and increase revenues, fog computing has recently been promoted by both industry and SDOs as one of the key components in 5G-supported IoT systems\cite{Chiang2017FogBook}. Compared to expensive massive-scale cloud data centers that are typically built in remote areas, fog computing consists of a large number of small computing servers, commonly referred to as fog nodes, that can offload computationally intensive tasks from the cloud data center to the edge of the network. Currently, major mobile network operators (MNOs) throughout the world are now actively upgrading their network infrastructure with fog nodes to provide extra-value-added services such as the IIoT and autonomous  vehicular services for their customers.

%Connecting a massive scale of IoT devices in 5G network with the support of network slicing and fog computing technology can significantly improve the reliability of data delivery and the performance of IoT services. One of the key challenges for supporting network slicing for massive IoT in a 5G network  is to quickly and efficiently allocate both networking and computational resources for serving different IoT services.

Recently, network slicing utilizing both communication and computational resources has attracted significant interest from both industry and academia. Allowing each slice to be supported by both types of resources can further improve the  performances of IoT services, balance resource utilization across different network elements, and open doorways for newly emerging IoT services with stringent latency and computational requirements. In spite of its great promise, allocating heterogenous  resources for multiple network slices with different service constraints introduces many novel challenges. First, different resources are typically managed by different service providers. Therefore, exchanging and sharing proprietary information such as resource availability and traffic dynamics between them are generally impossible. Second, both fog computing and communication network infrastructure as well as  IoT devices can be distributed throughout a wide geographical area and centralized coordination and management may result in intolerable coordination delay and excessive communication overhead. Finally, different types of IoT services with different features must be supported by various resources. How to design an optimal algorithm that can quickly and accurately allocate and isolate various combination of  resources \cite{caballero2019network} to support  network slices remains an open problem.

In this paper, we investigate the distributed network slicing for a 5G and fog computing- enabled IoT network consisting of a set of base stations (BSs) offering wireless communication services and a set of fog nodes performing computationally-intensive tasks close to the IoT devices. We consider the joint resource allocation of both bandwidth of BSs and processing power of  fog nodes for supporting multiple network slices at the same time. The main objective is to reduce the overall service latency experienced by IoT devices that typically include both communication delay in wireless links connecting IoT devices and BSs and the queuing delay at each fog node.
A novel distributed network slicing framework based on a new control plane element, federated-orchestrator (F-orchestrator), has been introduced.
In this framework, each BS can decide the resource allocation including both the bandwidth and computational resource for a set of requesting IoT services using its local data. The F-orchestrator can coordinate the resource allocation among all the BSs by requesting a single value, an intermediate result, from each BS. Our framework has a much lower communication overhead and can preserve the privacy of each BS. We propose a distributed algorithm based on Alternating Direction Method of Multipliers (ADMM) with Partial Variable Splitting (DistADMM-PVS) and discuss its implementation in our proposed network slicing framework. We prove that the proposed algorithm can achieve the global optimal resource allocation for a large IoT network with a linear convergence rate. To further reduce the coordination delay, we introduce an novel algorithm, referred to as asynchronized ADMM (AsyncADMM) to further improve the convergence performance without requiring perfectly synchronized coordination among BSs. We prove that the proposed algorithm is guaranteed to converge to the global optimal solution.  Finally, we consider a 5G and fog computing-supported smart transportation system as a case study to evaluate the performance of our proposed framework. The main contribution of this paper are summarized as follows:
%A distributed resource allocation algorithm has been proposed. We prove that the proposed algorithm can minimize the average latency of the entire network and at the same time guarantee satisfactory performance for each supported type of service. %In summary,
%
%Network slicing is introduced in the 5G SBA architecture. When different services have different requirements on resources and processing capabilities, different slices can be "customized" to support different services. At the same time, “customizing” the corresponding slices for different services can achieve a more reasonable allocation of resources under the premise of meeting the required delay of the service. Network slicing is very important. Some existing slicing methods are to slice the base station and antenna resources. When the tenant accesses the base station, it will involve the problem of bandwidth and processing resource allocation. Therefore, it is necessary to consider slicing some Other resources such as bandwidth and processing resources. At the same time, the existing slicing methods are considered to slice a single resource, and the current joint slicing is still difficult.
%
%This paper proposes a 5G SBA based bandwidth and processing resource joint slicing architecture.Existing slicing model usually considers a single resource, but actual scenarios often involve the allocation of multiple resources.Therefore,considering joint slicing is very meaningful.
%The main contributions of this paper are summarized as follows:
\begin{itemize}
\item[1)] A novel distributed network slicing framework, based on F-orchestrator, has been proposed. In this framework, each BS calculates the resource allocation based on its local information and only coordinate with other BSs via the F-orchestrator using intermediate calculation result. Each BS does not need to send its data information to other BSs nor F-orchestrator. Our framework can significantly reduce the communication overhead and preserve the privacy of each BS.
\item[2)] A distributed optimization algorithm, referred to as DistADMM-PVS, has been proposed to coordinate the resource allocation of both bandwidth of BSs and computational resources of fog nodes. We prove that our proposed algorithm can converge to the global optimal solution at the rate of $O(1/t)$.
    %with fast convergence rate. %  that can preserve the private information of both the base stations and fog nodes in a 5G network.
\item[3)] An improved algorithm, referred to as  AsyncADMM, has been proposed to reduce the coordination delay and further increase the convergence performance. We prove that the proposed algorithm can also converge to the global optimal solution.
\item[4)] We consider an IoT-based smart transportation system supported by a university campus 5G and fog computing networks as a case study. Extensive simulation has been conducted based on a set of road-side surveillance cameras recorded over two months of period. Our results show that our proposed distributed network slicing framework can offer significant  performance improvement in terms of service response time.
%Simulation and detailed performance analysis have been presented under various practical scenarios. Our result shows that joint slicing utilizing both bandwidth and computational resources offers around 15\% overall latency reduction compared to network slicing with only a single resource.
\end{itemize}

The rest of the paper is organized as follows. Related works are reviewed in Section \ref{Section_RelatedWork}. The distributed network slicing framework and F-orchestrator are introduced in Section \ref{Section_Architecture}. In Section \ref{Section_Model}, we present the system model and problem formulation.  Distributed optimization algorithms are presented in Section \ref{Section_Algorithm}. We evaluate the performance of our proposed framework by considering the campus smart transportation system as a case study in Section \ref{Section_SIMULATION}.  The paper is concluded in Section \ref{Section_Conclusion}. %Table~\ref{Table_Notation} presented the list of notations.

\vspace{-0.5cm}
 \section{Related Works}
\label{Section_RelatedWork}

{\bf Massive IoT networks}: Over the past decade, IoT systems have been extensively investigated by both industry and academia from various perspectives. In particular, 3GPP has introduced three IoT technologies for IoT systems: extended coverage-GSM  IoT  (EC-GSM-IoT),  narrowband-IoT  (NB-IoT),  and  enhanced  machine-type communication (eMTC) \cite{3GPP2016IoT}. Although these three technologies have different parameters and can provide different service performance with various resources, they are mostly focusing on low-cost low-power IoT services with limited performance guarantees \cite{Lin2017IoT}. ITU-T have significantly extended the application scenario and user cases of IoT \cite{ITU-T2016IoT} and consider IoT as a vision with technological and societal implications that can be used to offer services to all kinds of applications with diverse performance, security and privacy requirements. This sparked significant interest in both industry and academia to further investigate the architectures and  implementation details of IoT services with more stringent requirements. For example, in \cite{Dardari2016BroadbandIoT}, the authors have investigated the IoT for eMBB services. IoT  with applications requiring ultra-low latency and high reliability scenarios are investigated in\cite{Ren2019IoTURLLC}.

{\bf Network slicing with fog computing}:
Recently, network slicing utilizing computational resource in fog computing networks has attracted significant interest. In \cite{zhang2017computing}, the authors proposed a distributed optimization algorithm for the allocation of fog computing resources and applied it to improve the performance of IoT systems. In \cite{8437204}, the authors proposed a computational resource allocation scheme based on double-matching for fog computing networks. In\cite{Ni2018Efficient}, the authors  proposed an efficient and secure service-oriented authentication framework that can support both network slicing and fog computing for 5G-enabled IoT networks. In particular, the IoT users can efficiently establish their connections with 5G core network and  access IoT services through proper network slices of selected by fog nodes. In \cite{Xiao2018Dynamic}, the authors presented a dynamic network slicing framework  in which a regional orchestrator was introduced to coordinate workload distribution among local fog nodes, and  the amount of resources allocated to each slice can be dynamically adjusted according to service requests and energy availabilities.

{\bf Multi-resource allocation for network slicing}:
One of the main challenges for network slicing is how to quickly and effectively isolate and distribute multiple types of available resources according to the specific requirement of each service\cite{Ge2011CooperartiveCell}\cite{Ge2016UserMobility}\cite{Ge2015Energy}.
 In \cite{jiang2016network}, the authors studied the allocation of the radio resources for network slicing. A prioritized admission mechanism was proposed to improve the resource utilization and increase user's service experience. Network slicing has been studied   in with dynamic resource demand and availability mobile environment in \cite{zhang2017network}. In \cite{xiao2018distributed}, the authors proposed a network slicing architecture utilizing spectrum resources in both licensed and unlicensed bands. In\cite{Leconte2018Resource}, the authors propose an optimization framework that can jointly allocate resources for slices both in terms of network bandwidth and cloud processing powers in a fine-grained level. In this framework, the system designers are allowed to trade-off traffic-fairness with computing-fairness by tuning a slice-specific parameter.

\section{Distributed Network Slicing for IoT}
\label{Section_Architecture}

%\subsection{Architecture for 5G and Fog Computing  Enabled IoT Networks}
A generic network slicing architecture for supporting a diverse set of IoT services consists of the following elements:
\begin{enumerate}
\item {\it IoT tenants} correspond to over-the-top content or service providers that offer various smart services or contents based on the  data collected by IoT devices and processing result offered by  fog nodes. IoT tenants typically do not have to deploy any network infrastructure. They can however lease the networking and computational resources as well as data processing services provided by other providers.
\item {\it Wireless service providers (WSPs)} have carefully deployed their network infrastructure including both Base Stations (BSs) and the backhaul networks. IoT tenants will need to lease the networking resources for collecting and transporting the data collected by the IoT devices. If necessary, an IoT tenant can lease network infrastructure from multiple WSPs to further improve the service coverage and data delivery performance. %and connect IoT devices with their Base Stations (BSs) for data collection. An IoT network may tenant multiple WSPs for its data delivery.
\item {\it Fog computing service providers (FSPs)} deploy a large number of fog nodes close to the IoT devices to perform local data  processing services. Each fog node can be connected to BSs via local area network. %Two or more fog nodes can also exchange information among each other via intra-network connection or Internet.   %links and usually located closely to the IoT devices.
\item {\it IoT devices} are  IoT terminals each of which is associated with one specific type of IoT service with a certain QoS requirement. Each IoT device can connect to multiple BSs for data transmission.  %They are usually deployed throughout a wide geographical area.
\end{enumerate}

In this paper, we consider joint network slicing of both communication and computational resources for supporting a given set ${\cal N}$ of $N$ IoT services, labeled as ${\cal N} = \{1,2,...,N\}$.
Network slicing employs a network virtulization approach that virtualizes physical resources into Virtual Network Functions (VNFs). Each VNF can be further divided into smaller components to be placed in a common software container, so the network functionality can be quickly released and reused by different service instances. We refer to the smallest component that can be used in the VNFs for network slices as a NF. Each network slice can consist of many NFs. Different NFs are decoupled from each other. So each network slice can be launched and dynamically scaled without affecting other ongoing services.

%Network slicing is a virtualization framework in which the physical infrastructure including both communication and computational infrastructure as well as their attributes can be abstracted and logically partitioned into a set of virtualized network functions (NFs) each of which can be allocated to support a specific type of service. A {\em network slice} is a collection of these virtualized NFs to be isolated, controlled, and managed by the IoT tenants to deliver each specific type of service.

Since fog nodes need to be deployed close to the IoT devices, one of the ideal locations for deploying fog node is inside the BSs of the wireless access networks connecting the IoT devices. Therefore, most existing works assume fog nodes connected to BSs to minimize the latency for data delivery between IoT devicesand fog nodes. In this paper, we follow the same assumption and assume each fog node is connected with a BS and can serve  IoT devices located in the service coverage area. Each BS can support multiple IoT services allocating both its own bandwidth and computational resource of the fog node.

%\subsection{Existing network slicing architecture}

Currently, the most popular network slicing architecture is based on the OpenFlow architecture  introduced by ONF \cite{ONFSDNArchitecture}. In this architecture, an SDN controller instantiated in the control plane located between IoT tenants and hardware infrastructures  can monitor the availability of all the resource blocks and centrally control and allocate NF sets according to the requirements of the requesting services.

Although ONF's network slicing architecture offers a comprehensive solution for supporting various IoT services,  it cannot be directly applied to the massive-scale IoT network due to the following reasons:
\begin{enumerate}
\item {\bf High communication overhead}: Keeping the network status updated by all the physical network elements to the SDN controller and broadcasting the network slicing strategy to all the BSs and fog nodes will result in high communication overhead and excessive coordination delay that can be intolerable for some resouce-limit and delay-sensitive IoT services.
\item {\bf Limited flexibility and scalablity}: Each SDN controller also needs to create a complete abstract set to logically control the constitutes of every slice, which makes it difficult to support scalable and flexible network deployment especially in networks consists of a large number of devices.
\item {\bf Difficult to protect privacy}: In order to provide a proper resource allocation strategy for every IoT service, the SDN controller need to collect all the service requests from IoT devices and monitor resource usages and capacities of all the BSs and fog nodes which may not be possible for multi-operator supported scenarios. For example, in the case that multiple WSPs and FSPs jointly offer resources to support the same IoT service, it is generally impossible for all these service providers to  exchange and share proprietary information with each other.
\end{enumerate}

In this paper, we propose a communication-efficient federated network slicing framework \cite{Afolabi2018Survey} in which an orchestrator, referred to as federated-orchestrator (F-orchestrator), is deployed between tenants and physical infrastructure and service providers to perform distributed network slicing without requiring all their resource information. Our framework is inspired by the federated learning approach introduced in \cite{konevcny2016federated} in which a set of loose federations coordinated by a F-orchestrator. Each federation corresponds to a BS that can decide the resource usage and allocation based on the local information such as the local resource availability and locally received service request. In this case, each BS does not have to upload or disclose its local information to other service providers nor the F-orchestrator.

\begin{figure*}
  \centering
  \includegraphics[width=12cm]{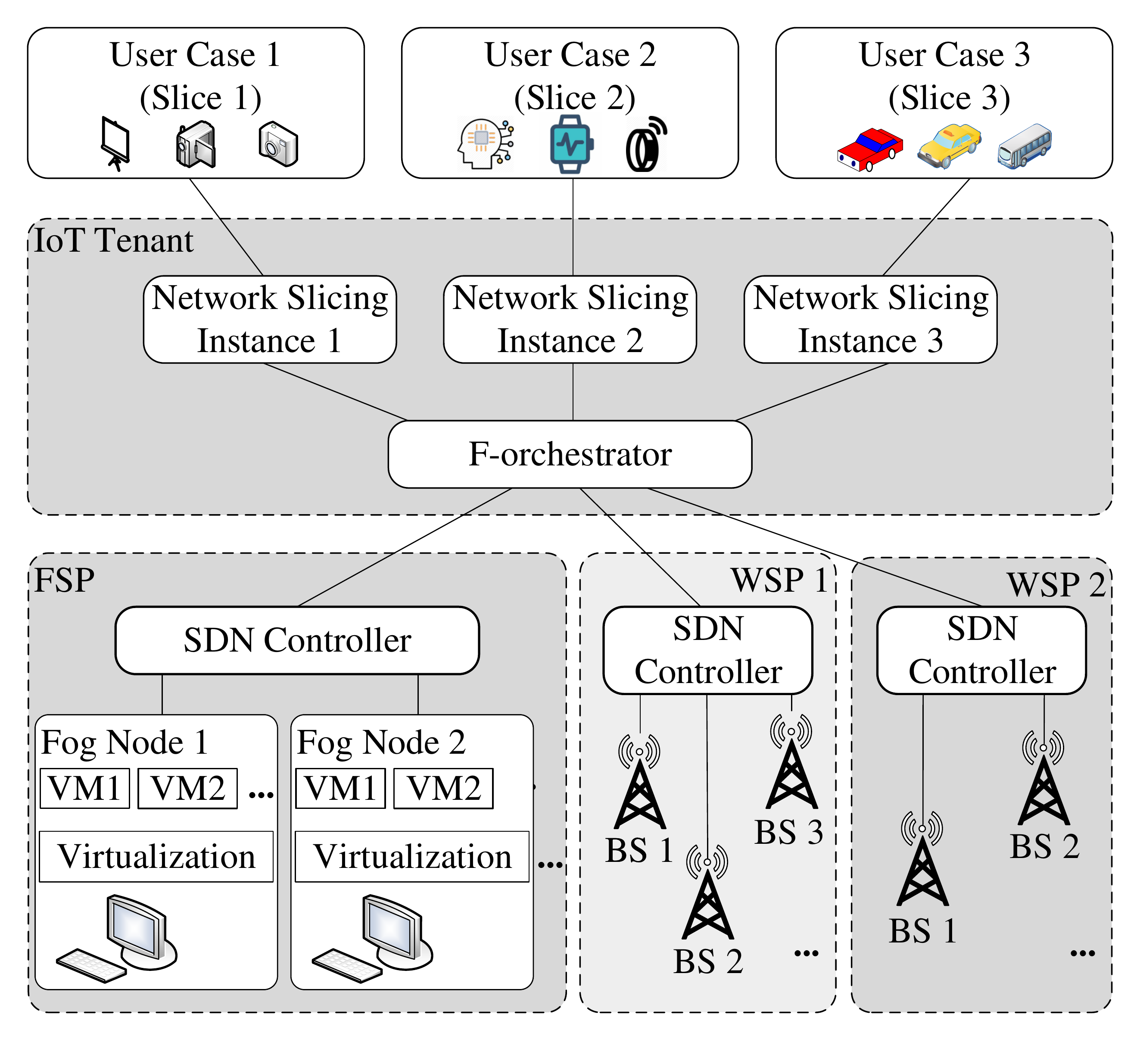}
  \vspace{-0.1in}
  \caption{5G network-slicing enabled IoT networks.}
  \begin{spacing}{-0.6}
%%行间距变为single-space
\end{spacing}
  \label{Architecture-1}
\end{figure*}

 Our proposed framework is illustrated in Figure~\ref{Architecture-1}. Under our distributed network slicing framework, each BS will decide the resource allocation according to the local service request and  resource availability information. The F-orchestrator will offer coordination services among all the BSs using their intermediate results to reach a global optimal resource allocation strategy for all the services requested by the IoT tenants.
In particular, our F-orchestrator has the following advantages:
\begin{enumerate}
\item {\bf Global optimal resource allocation}:  The F-orchestrator can coordinate all the BSs to achieve a global optimal resource allocation in spectrum and computational resources among all the BSs and fog nodes.% strategy as long as the QoS requirements of the IoT tenants can be formulated as a convex problem.
\item {\bf Privacy-preserving}: BSs do not need to share or exchange their proprietary information including resource availabilities and traffic dynamics with each other or with the F-orchestrator. The F-orchestrator only needs an intermediate result from (e.g., a dual variable as will be discussed in Section \ref{Section_Algorithm}) the each BS.
\item {\bf Communication-efficient}: Each BS only need to report one single intermediate result  to the F-orchestrator and get a feedback dual vector in each iteration. Compared to the centralized control approach with data monitoring and global resource management, the communication overhead in our proposed framework can be significantly reduced.
\item {\bf Support heterogeneous resources}: The F-orchestrator can coordinate resource allocation for heterogeneous resources. In this paper, we mainly focus on bandwidth and computational resource. Our proposed framework however can be easily extended to support other resources including energy consumption \cite{Ge2015EnergyEfficiency}\cite{Ge2013EnergyEfficiency} and caching capacities.
\end{enumerate}

%\subsection{Existing Network Slicing Architectures}
%\subsection{Proposed Architecture}

\section{System Model and Problem Formulation}
\label{Section_Model}

\begin{figure*}
  \centering
  \includegraphics[width=15cm]{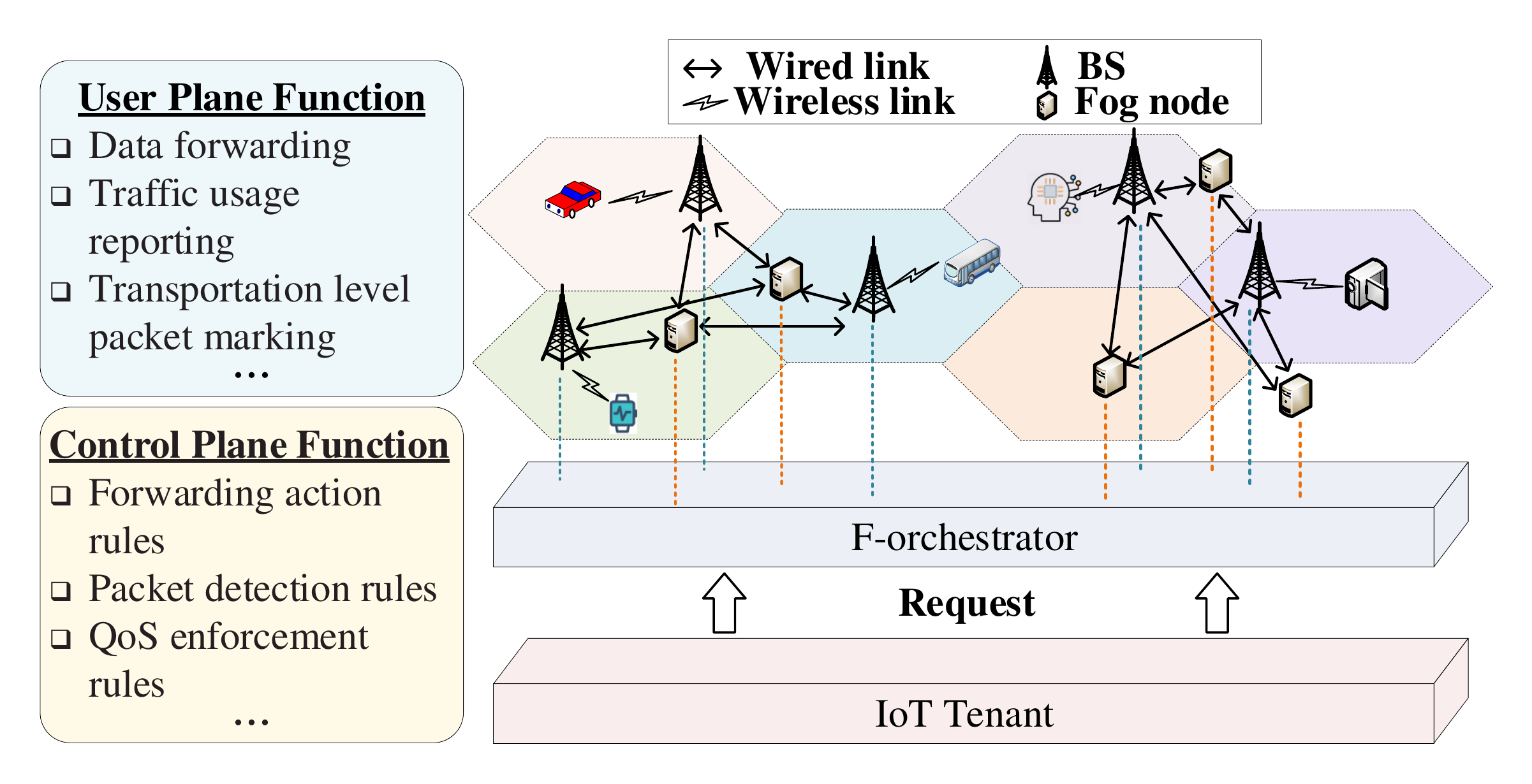}
  \vspace{-0.1in}
  \caption{Proposed distributed network slicing framework.}
  \begin{spacing}{-0.6}
%%行间距变为single-space
\end{spacing}
  \label{fig1}
\end{figure*}

\subsection{System Model}
We consider a network system consisting of a set $ \mathcal{F}=\{1\cdots F\}$ of $F$ fog nodes and a set $\mathcal{S}=\{1\cdots S\}$ of $S$ BSs as illustrated in Figure~\ref{fig1}. Each IoT device collect data associated with services in an exclusive coverage area. %Suppose each UE can request at most $N$ types of services. Let $\mathcal{N}=\{1\cdots N\}$ be the set of all supported types of services.
Each BS will query the F-orchestrator with the resource requests whenever it receives a service task . The F-orchestrator will then coordinate with the service requesting BS and connected fog nodes to create the corresponding network slices. The F-orchestrator will also supervise the resource reservation and routing of the service traffic between BSs and fog nodes. For each type of service, we assume that there exists a minimum volume of data, called {\it task unit}, that can be transmitted by BSs and to be processed by fog nodes. For example, in video or audio processing service, each video or audio clip consists of a number of video or audio data units for transporting and processing. Let $d_n$ be the data size of each task unit of service type $n$ for $n \in {\cal N}$. Each BS $s$ has been associated with a fixed bandwidth, labeled as $\beta_s$ for $s\in \mathcal{S}$, and each fog node $f$ can process at most $\mu_f$ task units per second for $f\in \mathcal{F}$. Suppose  arrival rate $k_{sn}$ of the $n$th service task unit at BS $s$ follows a Poisson distribution, $k_{sn}\sim P(\lambda_{sn})$, where $\lambda_{sn}$ is the expected number of received task units.

In this paper, we focus on minimizing the service response time for each type of service, which can consist of both {\it communication delay} for task unit transportation from IoT devices to BSs as well as the {\it queuing delay} at fog nodes. Let us first consider the communication delay. Note that in many practical networks, BSs and fog nodes can be connected with wireline or fiber which typically offers much higher data rate than the wireless links between IoT devices and BSs. Therefore, in this paper, we follow a commonly adopted setting and ignore the communication delay between BSs and fog nodes \cite{7239522}. For a given bandwidth $0\leq b_{sn}<\beta_s$ allocated by BS $s$ for service type $n$, we follow the commonly adopted setting \cite{6773024} and can write the communication delay for transporting each unit of task as
\begin{eqnarray}
p_{sn}={d_{n} \over {b_{sn} \cdot \log(1+h_{sn}\frac{w_{sn}}{\sigma_{sn}})}},
\label{communication delay}
\end{eqnarray}
where $w_{sn}$ is the transmission power to send each task units for service type $n$ from the IoT devices to  BS $s$, $\sigma_{sn}$ is the additive noise level received at BS $s$, and $h_{sn}$ is the channel gain between BS $s$ and the associated IoT device for service type $n$.

Queuing delay at the fog node can be affected by the processing power of fog nodes and task arrival rate. Suppose the maximum processing power allocated by fog nodes to process the $n$th type of service offered by BS $s$ for its associated IoT devices is $\mu_{sn}$ for $s\in \mathcal{S}$ and  $n \in {\cal N}$. We follow a commonly adopted setting and assume the task units processed by fog nodes can be modeled as M/M/1 queuing \cite{edelson1975congestion}. We can then write the queuing delay of the $n$th type of service in the coverage area of BS $s$ as
\begin{eqnarray}
q_{sn}={1 \over {\mu_{sn}-\lambda_{sn}}}.\label{queuing delay}
\end{eqnarray}

By combining (\ref{communication delay}) and (\ref{queuing delay}), the overall service response time of the $n$th type of service offered by BS $s$ can be written as
\vspace{-0.8cm}
\begin{eqnarray}
t_{sn}=p_{sn}+q_{sn}.
\end{eqnarray}
\vspace{-1.5cm}

\subsection{Problem Formulation}

In 3GPP's network slicing framework, a certain amount of resource must be reserved and isolated for each supported type of service, so there always exist available resources whenever a service request has been received and also the amount of the reserved resource must be higher than a certain threshold (e.g., sufficient to support each task unit). In this paper, we consider the F-orchestrator implemented in 3GPP's framework \cite{3GPPNetworkSlicing} \cite{3GPPManagement}. Let $b_0$ and $\lambda_{sn}$ be the minimized bandwidth and computational resources that mush be reserved in each time slot, respectively. %For a limited time duration, the F-orchestrator must first reserve a certain amount of computational resources at fog nodes, which is given by $\lambda_{sn}$, and a minimum bandwidth, which is given by  $b_0$ for task units of service type $n$ at BS $s$.
Thus, during each time slot,  we have the following constraints:
\begin{subequations}
\begin{align}
{b_{sn}} > {b_0} \quad\mbox{and}  \quad
{\mu _{sn}} > {\lambda _{sn}}, \quad \forall s \in \mathcal{S}, \; n \in \mathcal{N}
\end{align}
\vspace{-0.3in}
\end{subequations}

In this paper, we focus on the resource allocation and network slicing within a specific time duration in which the maximum bandwidth of a set of local BSs and a given amount of processing power of local fog nodes  reserved for a set of supported types of services can be considers as constants. The dynamic resources allocation and network slicing will be left for our future research.
We consider the following constraints:
\begin{itemize}
\item[1)] {\em Bandwidth constraint}: Let $\beta_s$ be the total bandwidth reserved by each BS $s$. In other words, the total bandwidth that can be allocated by BS $s$ to all upcoming service tasks cannot exceed $\beta_s$. Generally speaking, the F-orchestrator needs to reserve sufficient resources without knowing the exact number of task units which will be arrived in the future. It is however possible for the F-orchestrator to estimate the possible number of arrival task units according to the empirical probability distribution of the task arrival rate. In this way, the F-orchestrator can reserve sufficient resource to support the performance-guaranteed services for the majority of possible tasks with a certain level of confidence. More specifically, we define the confidence level $\theta=\Pr(k_{sn}\leq \theta_{sn})$ as the possibility that the number of type $n$ service task units arrived at BS $s$ is below a certain threshold number $\theta_{sn}$. For example, $\theta=0.9$ means that the F-orchestrator wants to reserve resources to meet the demands of all UEs with 90\% confidence. We can observe that $\theta$ is equivalent to the Cumulative Distribution Function (CDF) of task arrival rate $k_{sn}$. We can therefore write $\theta_{sn}=CDF_k^{-1}(\theta,\lambda_{sn})$ where $CDF_k^{-1}(\cdot)$ is the inverse function of CDF.
    We can then have the following constraint for the bandwidth allocated by BS $s$ to a set $\mathcal{N}$ of all supported types of services
\begin{eqnarray}
\sum _{n\in \mathcal{N}} \theta_{sn}\cdot b_{sn}\leq \beta_s.
\end{eqnarray}
\item[2)] {\em Computational resource constraint}: Suppose the total computational resource $\gamma$ reserved to all fog nodes  is limited. The summation of the computational resources allocated to all the services cannot exceed $\gamma$. We then have the following computational resources constraint
\begin{eqnarray}
\sum _{s\in \mathcal{S}}\sum _{n\in \mathcal{N}}\mu_{sn}\leq \gamma.
\end{eqnarray}
\end{itemize}

In addition, we assume each supported type of service $n$ has a maximum tolerable latency, labeled as $\overline {{T_{n}}}$, i.e., we have
\vspace{-0.8cm}
\begin{eqnarray}
{t_{sn}} \le \overline {{T_{n}}},\forall {s\in \cal S},{n \in \cal N}.
\end{eqnarray}

Our proposed framework is general and flexible. It can be applied to network slicing utilizing multiple resources across a wide geographical area.  In particular, we consider the following three network slicing cases.%In this paper, we consider network slicing architecture to leverage the RO.

\subsubsection{Bandwidth slicing}
In the bandwidth slicing case, only the spectrum resource allocation strategy for each type of service will be optimized, and all the computational resources are allocated linearly according to the proportion of $\lambda_{sn}$ , i.e.,  $\mu_{sn}=\gamma \cdot {\lambda_{sn} \over \sum\limits_{s=1}^{S}\sum\limits_{n=1}^{N}\lambda_{sn}}$.
The BSs are independent from each other and each of them only focuses on the allocation of bandwidth resources in its own coverage area. In this case, the service response time minimization problem can be formulated as:
\begin{subequations}
\begin{align}
\min_{\{b_{sn}\}_{s \in \mathcal{S},n \in \mathcal{N}}}  &\sum\limits_{s=1}^{S}\sum\limits_{n=1}^{N}t_{sn}\\
\mbox{ s.t. } \quad
&{b_{sn}} > {b_0},\forall s\in \cal S,\rm{n} \in \cal N,\\
&{t_{sn}} \le \overline {{T_{sn}}},\forall s\in \cal S,\rm{n} \in \cal N,\\
&\sum\limits_{n \in {\cal N}} {{\theta_{sn}}}  \cdot {b_{sn}} \le \beta_s,\forall s\in \cal S,\rm{n} \in \cal N.
\end{align}
\end{subequations}

%where $b_0$ is the minimum bandwidth resource allocated to the tasks of services. The bandwidth slicing architecture does not need to coordinate the BSs and the fog nodes, the optimization algorithm is relatively simple, which is beneficial to reduce communication overhead. However, during the entire communication process, the service response time is composed of communication delay and queuing delay, the bandwidth resources single slicing architecture only optimize the allocation of the bandwidth resources of the BSs, and the optimization problem of the computational resources is not considered. The fog node can only mechanically process the services transmitted by the BS, and cannot dynamically allocate the computational resources according to the task unit arrival rate of the service, which is easy to cause waste of computational resources and the overload of some fog nodes.
\subsubsection{Computational resources slicing}

In the computational resources slicing framework, only the processing power allocation strategy for each type of service will be optimized, and the bandwidth resources of BSs are allocated linearly according to the proportion of  data size of tasks in their coverage areas, i.e.,  $b_{sn}=\beta_s\cdot {d_n\cdot k_{sn} \over \sum\limits_{n=1}^{N} d_n\cdot k_{sn}}$. Therefore, the service response time minimization problem can be formulated as:
%The fog nodes or the fog nodes and BSs are independent of each other, each BS only focuses on the allocation of bandwidth resources in its own area. In summary, the problem we need to optimize is:
\begin{subequations}
\begin{align}
\min_{\{\mu_{sn}\}_{s \in \mathcal{S},n \in \mathcal{N}}}  &\sum\limits_{s=1}^{S}\sum\limits_{n=1}^{N}t_{sn}\\
\mbox{ s.t. } \quad
&{\mu _{sn}} > {\lambda _{sn}},\forall s\in \cal S,\rm{n} \in \cal N,\\
&{t_{sn}} \le \overline {{T_{sn}}},\forall s\in \cal S,\rm{n} \in \cal N,\\
&\sum\limits_{s \in {\cal S}} {\sum\limits_{n \in {\cal N}} {{\mu _{sn}}} }  \le \gamma,\forall s\in \cal S,\rm{n} \in \cal N.
\end{align}
\end{subequations}

%The computational resources slicing architecture has certain optimization capabilities for service response time, but the optimization effect is limited. Since the bandwidth resources of the BSs are not optimized, the allocation of bandwidth resources may be unfair, which causes the communication delay cannot be optimized, further degrading the performance of the entire network system.

\subsubsection{Joint network slicing}
In this paper, we consider network slicing architecture to leverage the F-orchestrator.
BSs and fog nodes can coordinate with the F-orchestrator to  allocate the bandwidth and the computational resources. In summary, we focus on designing a distributed algorithm to optimize the following problem
\begin{subequations}
\vspace{-0.1in}
\label{joint_prob}
\begin{align}
\min_{\{b_{sn}\}\{\mu_{sn}\}_{s \in \mathcal{S},n \in \mathcal{N}}}  &\sum _{s\in \mathcal{S}}\sum _{n\in \mathcal{N}}t_{sn}\\
\mbox{ s.t. } \quad
&{b_{sn}} > {b_0},\forall {s\in \cal S},{n \in \cal N},\\
&{\mu _{sn}} > {\lambda _{sn}},\forall {s\in \cal S},{n \in \cal N},\\
&{t_{sn}} \le \overline {{T_{n}}},\forall {s\in \cal S},{n \in \cal N},\\
&\sum\limits_{n \in {\cal N}} {{\theta_{sn}}}  \cdot {b_{sn}} \le \beta_s,\forall {s\in \cal S},{n \in \cal N},\\
&\sum\limits_{s \in {\cal S}} {\sum\limits_{n \in {\cal N}} {{\mu _{sn}}} }  \le \gamma,\forall {s\in \cal S},{n \in \cal N}.
\end{align}
\end{subequations}
%It can be seen from the above description that there are many parameters and constraints in our optimization problem, and there is virtually no information interworking between the BS and the fog nodes or between the BSs. In practice, it is difficult to implement with a centralized algorithm, so we consider using distributed algorithms.

\section{Distributed Optimization for Joint Network Slicing}
\label{Section_Algorithm}

%quickly and accurately allocate multiple resources supporting multiple network slices and
As mentioned before, in order to minimize the overall latency experienced by IoT devices, we need to carefully decide the amount of resources allocated to each network slice.
However, solving problem (\ref{joint_prob}) involves jointly deciding the proper amount of bandwidth and processing power for every type of services with global information such as the expected number of arrived task units and the computational capacity of every fog node which may result in intolerably high latency and communication overhead.

To address the above problems, we need to develop a  distributed optimization algorithm for solving the joint network slicing problem in  (\ref{joint_prob}) with the following design objectives:
\begin{enumerate}
  \item {\it Distributed Optimization with Coordination}: The proposed optimization algorithm should be able to separate the global optimization problem into a set of sub-problems, each of which can be solved by a BS according to its local information. The F-orchestrator can then be used to coordinate the solution of these subproblems to achieve the globally optimal resource allocation solution.
  \item {\it Privacy Preservation}: BSs and fog nodes may not want to share their private information such as bandwidths, expected number of arrived task units, and computational capacities with each other. %We would like to develop a framework that can preserve these information whiling solving the
  \item {\it Fast Convergence}: BSs and fog nodes connected to each F-orchestrator can change over the time. Thus, the algorithm that needs to quickly converge to the global optimal solution.
\end{enumerate}

In order to solve problem (\ref{joint_prob}) with the above objectives, we first propose a distributed Alternating Direction Method of Multipliers (ADMM) algorithm with Partial Variable Splitting, referred to as DistADMM-PVS. In this algorithm,  the service reponse time minimization problem is divided to a set of sub-problems, each of which is solved by a BS according to its local private information. The F-orchestrator will perform a synchronized coordination for the BSs after collecting all their intermediate results, and it will then broadcast the updated dual variable to all the BSs to start a new round of iteration. Such an synchronized approach can preserve the private information of all the network elements  and quickly converge to a global optimal value, but will also introduce an intolerable  synchronization delay especially for massive-scale networking systems. To solve this problem,  we then propose an  asynchronous ADMM-based distributed optimization algorithm, referred to as AsyncADMM, in which each BS can get immediate feedbacks from the  F-orchestrator after submitting its intermediate result.
\vspace{-0.6cm}
\subsection{Distributed ADMM Algorithm with PVS}

In this subsection, we propose a distributed optimization algorithm based on ADMM. Compared to traditional convex optimization algorithms, ADMM is more suitable for solving inequality constrained optimization problems in a decentralized manner. Furthermore, the decomposition-coordination procedure of the ADMM makes it possible to protect the aforementioned private information of BSs and fog nodes. Unfortunately, normal ADMM approaches can only handle problems with two blocks of variables \cite{boyd2011distributed}. In this algorithm, the Lagrangian dual problem of  (\ref{joint_prob}) will be divided to $S$ sub-problems, each of which can be solved by an individual BS using its local information. The F-orchestrator will collect the intermediate results from BSs and send the coordination feedbacks.    %coordinate between these sub-problems to achieve global optima

Let us first follow the same line as \cite{boyd2011distributed} and combine constraints in (\ref{joint_prob}) with the objective function by introducing a set of $S+1$ indicator functions. Specifically, for constraints (\ref{joint_prob}b)-(\ref{joint_prob}e) that can be separated across different BSs, we define $\cG_s = \{\bb_{s  }, \bmu_{s  } : {b_{sn}} > {b_0}, {\mu _{sn}} > {\lambda _{sn}}, {t_{sn}} \le \overline {{T_{n}}}, \sum_{n \in {\cal N}} {{\theta_{sn}}}\cdot {b_{sn}} \le \beta_s,\forall {s\in \cal S}, {n\in \cal N}\}$ as the feasible set corresponding to BS $s$ where $\bb_{s  } = \langle b_{sn} \rangle_{n \in \cal N}$ is the vector of bandwidth allocated by BS $s$ for each type of services and $\bmu_{s  } = \langle \mu_{sn} \rangle_{n \in \cal N}$ is the vector of processing power allocated for each type of services connected to BS $s$. Let $\bx_{s  } =\langle \bb_{s  }, \bmu_{s  } \rangle, \forall s\in \cal S$, we define $S$ indicator functions as
\begin{eqnarray}
{\bf I}_{\cG_s} \left(\bx_{s  } \right) = \left\{ {\begin{array}{*{20}{c}}
{0,} & \bx_{s  } \in \cG_s, \\
{+\infty,} & \bx_{s  } \notin \cG_s,
\end{array}} \right. \quad \forall s\in \cal S.
\end{eqnarray}

For constraint (\ref{joint_prob}f) that cannot be separated, we can also define an indicator function ${\bf I}_{\cG} \left(\bmu \right)$ as
 \begin{eqnarray}
{\bf I}_{\cG} \left(\bmu \right) = \left\{ {\begin{array}{*{20}{c}}
{0,} & \bmu \in \cG, \\
{+\infty,} & \bmu \notin \cG,
\end{array}} \right.
\end{eqnarray}
 where $\bmu = [\bmu_1, \bmu_2, \ldots, \bmu_S]$, and $\cG$ is the half space defined by $\cG = \{\bmu: \sum_{s \in {\cal S}} {\sum_{n \in {\cal N}} {{\mu _{sn}}} }  \le \gamma\} $.

By including the above indicator functions ${\bf I}_{\cG_s}$ and ${\bf I}_{\cG} $, the original joint network slicing problem (\ref{joint_prob}) with a set of inequality constraints can be converted to the following form without inequality constraints
\begin{subequations} \label{Sep_Prob}
  \begin{align}
    \min_{\{\bx_{s  }, \bz_{s  }\}_{s\in {\cal S}}} & \quad \sum_{s\in {\cal S}} \left\{ f(\bx_{s }) + {\bf I}_{\cG_s} \left(\bx_{s} \right) \right\} + \bf I_{\cG} \left(\bz \right)      \label{Sep_Obj}             \\
    \mbox{ s.t. }    & \quad \bx_{s}= \bz _{s} , \forall s \in {\cal S},
  \end{align}
\end{subequations}
where $f(\bx_{s }) = \sum_{n\in \cal N} t_{sn}$, and $\bz= [\bz_1, \bz_2, \ldots, \bz_S]$ is the introduced auxiliary variable.

The augmented Lagrangian of problem (\ref{Sep_Prob}) is given by
\begin{eqnarray}
  {\cal L}_{\rho} (\bx, \bz, \bLambda) = \sum_{s \in {\cal S}} \left\{ f(\bx_{s}) + {\bf I}_{\cG_s} \left(\bx_{s} \right) \right\} + \bf I_{\cG} \left(\bz \right)    + \bLambda^T(\bx-\bz) + \frac{\rho}{2} ||\bx-\bz||_2^2 ,
   \label{Lag}
\end{eqnarray}
where $\bLambda$ is the dual variable and $\rho$ is the augmented Lagrangian parameter.

\begin{theorem}
\label{Theorem_Convex}
The augmented Lagrangian of problem (\ref{Sep_Prob}) specified in (\ref{Lag})  is convex and partially separable among $\bx_s$.
\end{theorem}
\begin{IEEEproof}
Please refer to Appendix~\ref{Theorem_1} for the detailed proof.
\end{IEEEproof}

Therefore, we can then convert problem (\ref{Sep_Prob}) into two-block ADMM form as follows
\begin{subequations}
\begin{align}
{\bx^{k + 1}} & = \mathrm{\mathop {argmin}\limits_{\bx}} \sum_{s\in {\cal S}} \left\{ f(\bx_{s }) + {\bf I}_{\cG_s} \left(\bx_{s} \right) \right\}
+ \frac{\rho }{2}\|\bx - \bz^{k}+\bLambda^{k}\|_2^2  \label{CADMM_x},\\
{\bz^{k + 1}} &= \mathrm{\mathop {argmin}\limits_{\bz}} \bf I_{\cG} \left(\bz \right)  + \frac{\rho }{2}\|\bx^{k+1} - \bz+\bLambda^{k}\|_2^2 \label{CADMM_z},\\
{\bLambda^{k + 1}} &=\bLambda^{k}+\bx^{k+1}-\bz^{k+1},\label{CADMM_lambda}
\end{align}
\end{subequations}
where $k$ denotes the number of iterations.  According to the partially separability of $\cal L_{\rho} (\bx, \bz, \bLambda) $,  we can divide (\ref{CADMM_x}) into a set of sub-problems, each of which can be solved by a BS using its local information. In particular, each BS $s$ solves the following sub-problem
\begin{eqnarray}
   \bx_s^{k + 1} &=& \mathrm{\mathop {argmin}\limits_{\bx_s}} f(\bx_{s }) + {\bf I}_{\cG_s} \left(\bx_{s} \right) + \frac{\rho }{2}\|\bx_s - \bz_s^{k}+\bLambda_s^{k}\|_2^2, \quad \forall s \in \cal S.\label{DistADMM_x}
\end{eqnarray}

Meanwhile, (\ref{CADMM_z}) is equivalent to projecting the point $\bx^{k+1}+\bLambda^{k}$ onto the halfspace $\cG$, i.e.
\begin{equation}\label{DistADMM_z}
  {\bz^{k + 1}} = {\bPi_{\cG}}(\bx^{k+1}+\bLambda^{k}),
\end{equation}
where $ {\bPi_{\cG}}(\cdot)$ denotes the projection onto halfspace $\cG$.

Detailed description of our proposed algorithm is presented in Algorithm 1.

\begin{theorem}
\label{Theorem_2}
The proposed DistADMM-PVS algorithm converges to the global optimal solution of the joint network slicing problem in (\ref{joint_prob}) at a rate of $O(1/t)$.
\end{theorem}
\begin{IEEEproof}
Since the distributed sub-problems specified in (\ref{DistADMM_x}) is equivalent to the centralized $\bx$-update in (\ref{CADMM_x}), the convergence property of our proposed DistADMM-PVS algorithm directly follows that of the standard ADMM approach \cite{boyd2011distributed}. For the details please refer to Appendix~\ref{AppendixA}
\end{IEEEproof}

\begin{algorithm}
\footnotesize
  \caption{Distributed ADMM with Partial Variable Splitting (DistADMM-PVS)}\label{Algorithm 1}
  \begin{itemize}
    \item[] Initialization: Each BS $s$ chooses an initial variable $\bx_s^0$ and the F-orchestrator chooses an initial dual variable $\bLambda^0$; $k=1$
    \item[] Set the maximum number of iterations as $K>0$
    \item[] {\bf while} $k \leq K$ {\bf do}
          \begin{itemize}
            \item[] 1. Each BS $s$ simultaneously do:
                  \begin{itemize}
                    \item[] 1) Update $\bx_s^{k+1}$ according to (\ref{DistADMM_x}) and report it to the F-orchestrator;
                    \item[] 2) Allocate its bandwidth for each type of arrived task according to $\bb_s^{k+1}$;
                  \end{itemize}
            \item[] 2. After all the $\bx_s^{k+1}$ are received, the F-orchestrator do:
                  %\begin{itemize}
                  \begin{itemize}
                    \item[] 1) Update auxiliary variable $\bz^{k+1}$ according to (\ref{DistADMM_z});
                    \item[] 2) Update dual variable $\bLambda^{k+1}$ according to (\ref{CADMM_lambda});
                    \item[] 3) {\bf if} {Stopping criteria met}
                    \item[]  \quad\quad break;
                    \item[] \quad {\bf end if}
                    \item[] 4) Sends the sub-vectors $\bz_s^{k+1}$ and $\bLambda_s^{k+1}$ to the corresponding BS $s$;
                  \end{itemize}
            \item[] 3. $k=k+1$
          \end{itemize}
    \item[] {\bf end while}
  \end{itemize}
\end{algorithm}

In  our proposed DistADMM-PVS algorithm, each BS calculate its own subproblem, when all BSs finish calculating, the total results are sent to F-orchestrator, and F-orchestrator will update dual variable. As mentioned earlier, all the BSs must wait at idle until the slowest one among them finished its subproblem, which will cause intolerable  synchronization delays as show in Figure~\ref{sync} and may even cripple the whole system.

%\begin{figure}
%\centering
%\includegraphics[width=9cm]{synchronous_update.pdf}
%\caption{Synchronous Update of BSs}
%\label{synchronous}
%\end{figure}
%\begin{figure}
%\centering
%\includegraphics[width=9cm]{asynchronous_update.pdf}
%\caption{Asynchronous Update of BSs}
%\label{asynchronous}
%\end{figure}

%\section{}
%\label{Section_asynchronousAlgorithm}
\vspace{-0.5cm}
\subsection{Distributed Asynchronized ADMM Algorithm}

\begin{figure}
 \subfigure[Synchronous Updates of BSs]{
\begin{minipage}[t]{0.5\linewidth}
\centering
\includegraphics[width=8cm]{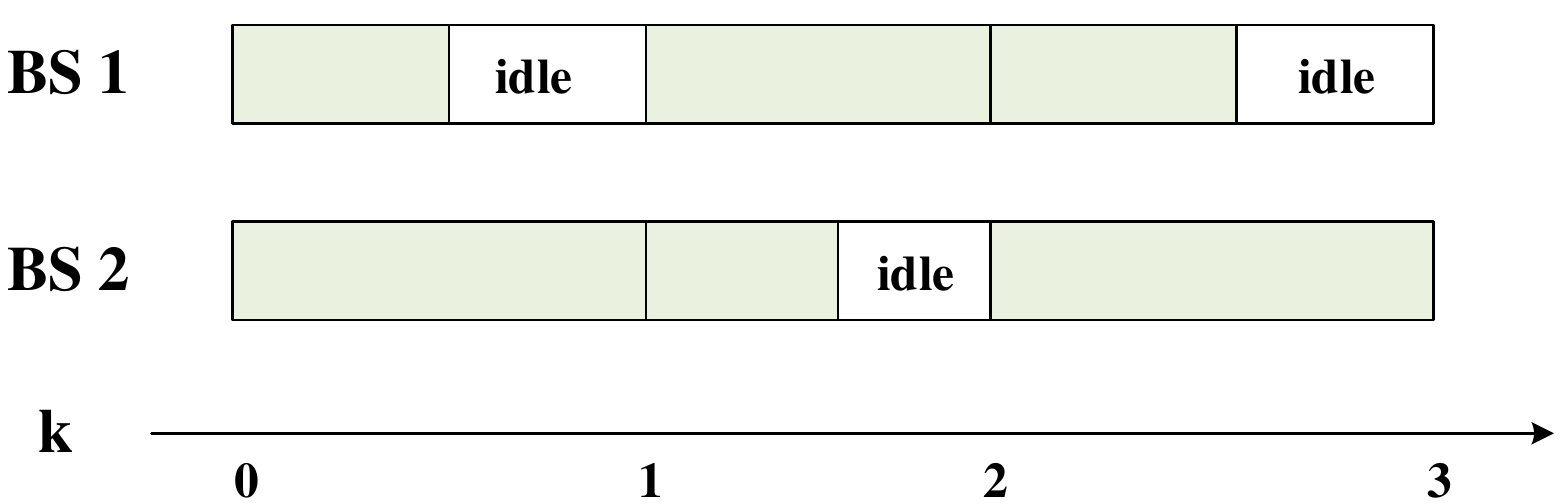}
%\vspace{-0.3in}
%\caption{PDF of H and V Service in area 9}
\label{sync}
\end{minipage}
}
\subfigure[Asynchronous Updates of BSs]{
\begin{minipage}[t]{0.5\linewidth}
\centering
\includegraphics[width=8cm]{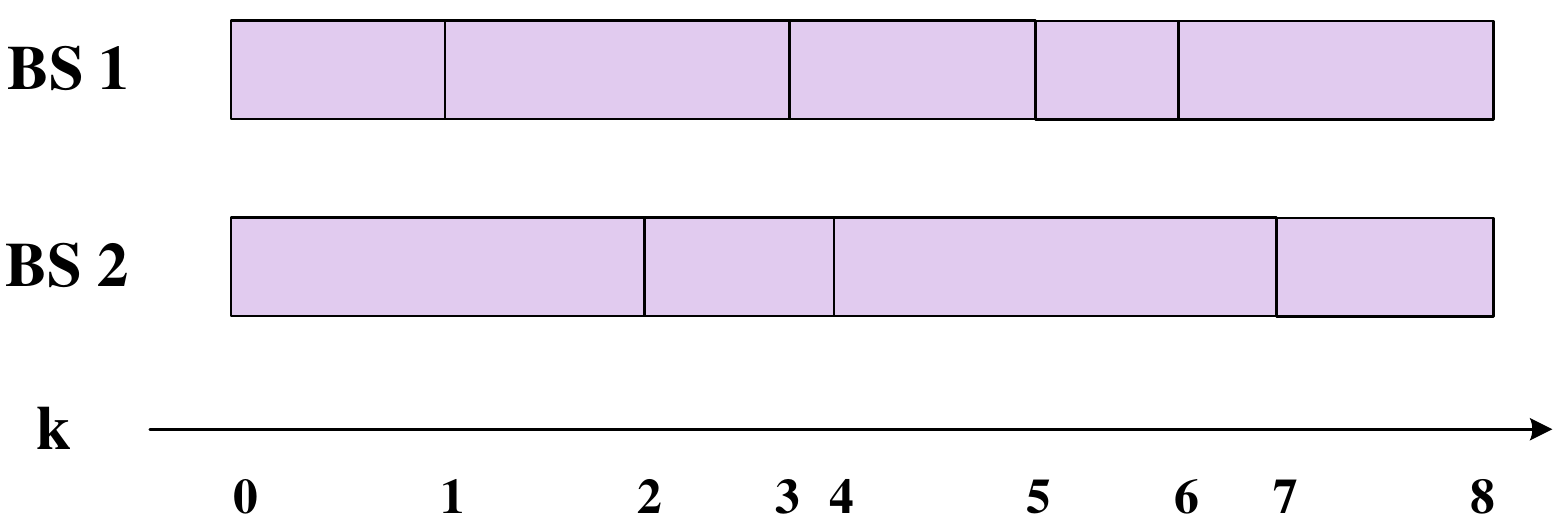}
%\vspace{-0.3in}
%\caption{PDF of H and V Service of all areas}
\label{async}
\end{minipage}
}
\caption{Synchronized updates versus asynchronized updates. In the synchronized approach, all the BSs must wait at idle until the slowest one among them finished its subproblem, which will cause intolerable synchronization delay.}
\vspace{-0.5cm}
\end{figure}

\begin{figure}
\centering
\includegraphics[width=9cm]{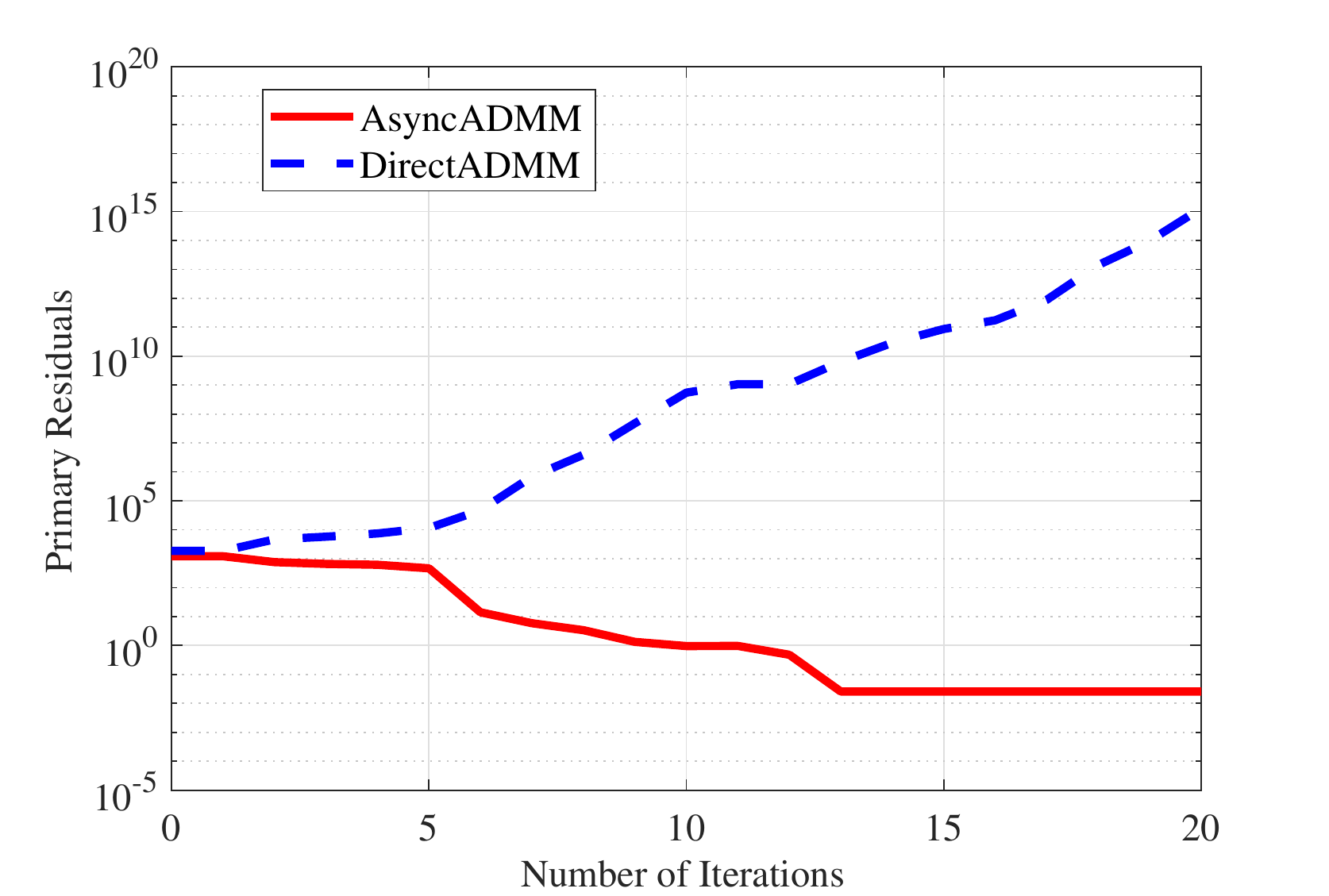}
%\vspace{-0.3in}
\caption{Convergence performance of the directly asynchronized approach. The  primal residual keeps increasing instead of decreasing as expected. In contrast, our proposed AsyncADMM approach can converge. }
\label{fig_DirectADMM}
\vspace{-0.5cm}
\end{figure}

To solve the aforementioned problem in synchronized algorithms,  we propose an asynchronous ADMM-based distributed optimization algorithm that can avoid the idle waiting of BSs. In this algorithm, the F-orchestrator performs a dual update immediately after it receives an update from a single BS. As s result, there will be
no synchronization delay and the global optimal resource allocation can be achieved more efficiently. Besides,  the system will be more tolerant to computing faults and communication glitches and easy to incorporate more BSs \cite{peng2016arock}. However, directly perform asynchronous updates in a distributed algorithm may fail to converge  due to the following reasons:
\begin{enumerate}
  \item The subproblems maybe coupled from each other. From (\ref{CADMM_x})-(\ref{CADMM_lambda}), we can see that the $\bx$-subproblem and $\bLambda$-subproblem can be completely separated, i.e., each BS $s$ can calculate $\bx_s^{k + 1}$ and $\bLambda_s^{k + 1}$ without the intermediate results from the $k$th iteration . However, the $\bz$-problem cannot be divided into separated subproblems. In this case, in order to update $\bz^{k + 1}$, we need to know collect all the $\bz_s^{k }$ for $s \in \mathcal{S}$. As a result, Algorithm 1 can not be directly performed in a asynchronized manner.
  \item It is more difficult to analyze asynchronous algorithms and ensure their convergence.  For example, in Figure \ref{async}, at the 6th iteration, BS 1 has been updated 4 times while BS 2 only finishes its 2nd update. Therefor, BS 1 can only receive coordination result from the F-orchestrator based on outdated information of BS 2. In this case, it becomes impossible to find a sequence of iterates that one completely determines the next, and the algorithm may fail to converge, as illustrated in Figure~\ref{fig_DirectADMM}.
\end{enumerate}

%\begin{figure*}
%\begin{minipage}[t]{0.5\linewidth}
%\centering
%\includegraphics[width=7.5cm]{synchronous_update.pdf}
%\caption*{Synchronous Updates of BSs}
%\label{sync}
%\end{minipage}
%\begin{minipage}[t]{0.5\linewidth}
%\centering
%\includegraphics[width=7.5cm]{asynchronous_update.pdf}
%\caption*{Asynchronous Updates of BSs}
%\label{async}
%\end{minipage}
%%\caption{Synchronized updates versus asynchronized updates. In the synchronized approach, all the BSs must wait at idle until the slowest one among them finished its subproblem, which will cause intolerable synchronization delay.}
%%\label{delay}
%\end{figure*}

%\begin{figure}
%\centering
%\includegraphics[width=9cm]{synchronous_update.pdf}
%\caption{Synchronous Updates of BSs}
%\label{sync}
%\end{figure}
%\begin{figure}
%\centering
%\includegraphics[width=9cm]{asynchronous_update.pdf}
%\caption{Asynchronous Updates of BSs}
%\label{async}
%\end{figure}

In \cite{peng2016arock}, the authors proposed a asynchronous parallel algorithm and the generated sequence is guaranteed to converge to the global optimal solution, even with the existence with outdated updates from other agents. In this paper, we extend this  asynchronous parallel computing
scheme in \cite{peng2016arock}\cite{peng2016coordinate} to our proposed distributed network slicing framework with F-orchestrator and
propose an AsyncADMM approach.

\begin{comment}
\begin{align}
x^{*} = \mathcal{T}x^{*},\quad for \quad  x^{*} \in \mathcal{H},\label{fixed_point}
\end{align}

where $\mathcal{H}$ means Hilbert space. The authors adopted the $Krasnosel'ski\breve{i}-Mann(KM)$ iteration to solve problem (\ref{fixed_point}) as follows
\begin{subequations}
\begin{align}
x^{k+1} = x^{k}+\alpha(\mathcal{T}x^{k}-x^{k}),\label{KM_a} \\
\mathrm{\mathop {equivalently}}, x^{k+1} = x^{k}-\alpha\mathcal{S}x^{k},\label{KM_b}
\end{align}
\end{subequations}
where $\mathcal{S}\equiv I-\mathcal{T}$.
\end{comment}

In order to solve problem (\ref{Sep_Prob}) in an asynchronized manner, we consider its dual form in which the subproblems can be completely separated across all the BSs. Let $\bF_s(\bx)= f(\bx_{s }) + {\bf I}_{\cG_s} \left(\bx_{s} \right) $, and $\bF(\bx)= \sum_{s\in {\cal S}} \left\{ f(\bx_{s }) + {\bf I}_{\cG_s} \left(\bx_{s} \right) \right\}$, then the dual problem of (\ref{Sep_Prob})  is
\begin{subequations} \label{Dual_Prob}
  \begin{align}
    {\mathrm{\mathop{min}\limits_{\bLambda}}}  \quad  \bF^{*}(\bLambda)+ \bf I^{*}_{\cG} \left(\bLambda \right),
  \end{align}
\end{subequations}
where $\bF^{*}(\bLambda)= \max (\bLambda^{T}\bx-\bF^{*}(\bx))$ and $\bf I^{*}_{\cG} \left(\bLambda \right)= \max (-\bLambda^{T}\bz-\bf I^{*}_{\cG} \left(\bz \right))$.

As mentioned in \cite{eckstein1992douglas}, the ADMM algorithm is equivalent to the Douglas-Rachford splitting method(DRSM). In this paper, we can define the DRSM operator of problem (\ref{Dual_Prob})  as:
\begin{align}
%\mathcal{M}_{DRS}^{k+1}={1\over 2}[\mathcal{M}_{DRS}^{k}+(2{\bf prox}_{\rho\bF^{*}}-I)(2{\bf prox}_{\rho\bf I^{*}_{\cG}}-I)(\mathcal{M}_{DRS}^{k})],
\mathcal{M}_{DRS}={1\over 2}[\mathcal{I}+(2{\bf prox}_{\rho\bF^{*}}-\mathcal{I})(2{\bf prox}_{\rho\bf I^{*}_{\cG}}-\mathcal{I})],
\end{align}
where $\mathcal{I}$ is the identity operator, $\bf prox$ is the proximal operator, and we can also define $\mathcal{S}_{DRS}$ is as follows
\begin{eqnarray}
    \mathcal{S}_{DRS}&=&\mathcal{I} - \mathcal{M}_{DRS}\nonumber\\
    &=&{1\over 2}[\mathcal{I}-(2{\bf prox}_{\rho\bF^{*}}-\mathcal{I})(2{\bf prox}_{\rho\bf I^{*}_{\cG}}-\mathcal{I})]\label{S_DRS}
\end{eqnarray}

According to \cite{peng2016arock}, the above problem (\ref{Dual_Prob}) is a fixed point problem that can be solved in an asynchronous parallel manner via:
\begin{eqnarray}\label{KM}
\bV^{k+1} &=&\bV^{k}-\alpha_k\mathcal{S}\hat{\bV^{k}}\nonumber\\
&=&\bV^{k}-\alpha_k(\mathcal{U}_{s_k}\circ\mathcal{S})\hat{x^{k}},
\end{eqnarray}
where $\bV$ is the introduced temporal variable that satisfies $\bLambda={\bf prox}_{\rho\bf I^{*}_{\cG}}(\bV)$,  $s_k$ is the activated BS at the $k$-th iteration, $\mathcal{U}_{s_k}$ is the sub-vector selection operator, and $\mathcal{U}_{s_k}\circ\mathcal{S}$ means in $k$-th iteration, only the $s_k$th-subproblem is updated. $\hat{x^{k}}$ is the feedback vector that BS $s_k$ received from the F-orchestrator during the $k$-th iteration, and $\alpha_k$ is the stepsize. %, and $x^{k}$ is the updating result of the last iteration. %The authors proved that the Arock asynchronous algorithm proposed in \cite{peng2016arock} can converge to a optimal value, in this paper, we consider to extend the Arock asynchronous algorithm to a distributed asynchronous ADMM algorithm and apply it to our problem (\ref{Sep_Prob}).
In this way,  problem (\ref{Dual_Prob}) can be solved via:
\begin{subequations}
\label{async1}
\begin{align}
\mathbf{Itreation}:\quad &\bV^{k+1}=\bV^{k}-\alpha^{k}(\mathcal{U}_{s_k}\circ\mathcal{S}_{DRS})\hat{\bV}^{k},\label{asynchronous_update}\\
\mathbf{Return}:\quad &\bLambda^{k+1}={\bf prox}_{\rho\bf I^{*}_{\cG}}(\bV^{k+1}).
\end{align}
\end{subequations}
%
%We consider to transform (\ref{V_Itreation}) to a asynchronous update form, which is as follows
%\begin{align}
%\bV^{k+1}=\bV^{k}-\alpha^{k}(\mathbf{U}_{s_k}\circ\mathcal{S}_{DRS})\hat{\bV}^{k},\label{asynchronous_update}
%\end{align}
%where $(\mathcal{U}_{s_k}\circ\mathcal{S}_{DRS})\hat{\bV}^{k}$ means that in kth iteration only BS $s_k$ will be updated,
To be more specific, the asynchronous update step (\ref{asynchronous_update}) can be calculated through
\begin{eqnarray}
\bV_s^{k+1} = \left\{ {\begin{array}{*{20}{c}}
{\bV_s^{k}-\alpha^{k}\mathcal{S}_{DRS}\hat{\bV_s^{k}},} & s=s_k, \\
{\bV_s^{k},} & s\neq s_k,
\end{array}} \right.\quad \forall s\in \cal S
\label{asynchronous_updatenew}
\end{eqnarray}

After plugging the definition of $\mathcal{S}_{DRS}$ as shown in (\ref{S_DRS}) into the above updated in (\ref{async1}) and  (\ref{asynchronous_updatenew}), we can get the following results:
\begin{subequations}
\label{async2}
  \begin{align}
   & \bV^{k+1}=\bV^{k}-\alpha^{k}\mathcal{U}_{s_k}(\hat{\bLambda}_{\rho\bf I^{*}_{\cG}}^{k}-\hat{\bLambda}_{\rho\bF^{*}}^{k})\label{v_k+1}\\
    &\hat{\bLambda}_{\rho\bf I^{*}_{\cG}}^{k}= {\bf prox}_{\rho\bf I^{*}_{\cG}}(\hat{\bV^{k}}),\label{lambda_Ig}\\
      &\hat{\bLambda}_{\rho\bF^{*}}^{k}= {\bf prox}_{\rho\bF^{*}}(2\hat{\bLambda}_{\rho\bf I^{*}_{\cG}}^{k}-\hat{\bV^{k}}).\label{lambda_F}
  \end{align}
\end{subequations}

To derive a more detailed version of the updates in (\ref{async2}), we can reformulate (\ref{lambda_Ig}) as
\begin{subequations}
\begin{align}
\hat{\bLambda}_{\rho\bf I^{*}_{\cG}}^{k}=&\mathrm{\mathop {argmin}\limits_{\bLambda}} [ I^{*}_{\cG} \left(\bLambda \right)  + \frac{1 }{2\rho}\|\bLambda-\hat{\bV^{k}}\|_2^2]\\
=& \mathrm{\mathop {min}\limits_{\bLambda}} [\mathrm{\mathop {max}\limits_{\bz}} (-\bf I_{\cG} \left(\bz \right)-\bLambda^{T}\bz)  + \frac{1 }{2\rho}\|\bLambda-\hat{\bV^{k}}\|_2^2]\label{max_z}\\
=& \mathrm{\mathop {max}\limits_{\bz}}  [\mathrm{\mathop {min}\limits_{\bLambda}}  (-\bf I_{\cG} \left(\bz \right)-\bLambda^{T}\bz)  + \frac{1 }{2\rho}\|\bLambda-\hat{\bV^{k}}\|_2^2]\label{min_lambda}.
\end{align}
\end{subequations}
Here, the optimal solution of $\bLambda$ can be derived as
\begin{subequations}
\begin{align}
\bLambda^{*}=&\mathrm{\mathop {argmin}\limits_{\bLambda}}  (-\bf I_{\cG} \left(\bz \right)-\bLambda^{T}\bz)  + \frac{1 }{2\rho}\|\bLambda-\hat{\bV^{k}}\|_2^2\\
=& \hat{\bV^{k}}+\rho\bz\label{Lambda_solution},
\vspace{-0.5cm}
\end{align}

\end{subequations}
\vspace{-0.5cm}
and if we plug (\ref{Lambda_solution}) back to (\ref{min_lambda}), we can then derive the  optimal solution of $\bz$ in (\ref{max_z}) as
\begin{subequations}
\begin{align}
\bz^{*}=&\mathrm{\mathop {max}\limits_{\bz}}  -\bf I_{\cG} \left(\bz \right)-(\hat{\bV^{k}}+\rho\bz)^{T}\bz  + \frac{\rho }{2}\|\bz\|_2^2\\
=&\mathrm{\mathop {max}\limits_{\bz}}  -\bf I_{\cG} \left(\bz \right)-(\hat{\bV}^{k})^{T}\bz  - \frac{\rho }{2}\|\bz\|_2^2\\
=&\mathrm{\mathop {argmin}\limits_{\bz}}  \bf I_{\cG} \left(\bz \right)+\bz^{T}\hat{\bV^{k}}  + \frac{\rho }{2}\|\bz\|_2^2 \label{z_solution}.
\end{align}
\end{subequations}
Combing (\ref{z_solution}) and (\ref{Lambda_solution}), the update in (\ref{lambda_Ig}) can be performed via
\begin{eqnarray}\label{z_k}
\left\{ {\begin{array}{*{20}{c}}
{\hat{\bz}^{k}=\mathrm{\mathop {argmin}\limits_{\bz}}  \bf I_{\cG} \left(\bz \right)+\bz^{T}\hat{\bV^{k}}+ \frac{\rho }{2}\|\bz\|_2^2,}  \\
{\hat{\bLambda}_{\rho\bf I^{*}_{\cG}}^{k}=\hat{\bV^{k}}+\rho\hat{\bz}^{k}} .
\end{array}}\right.
\end{eqnarray}

Following the same line, we can also derive the detailed steps for (\ref{lambda_F}) as follows
\begin{eqnarray}\label{x_k}
\left\{ {\begin{array}{*{20}{c}}
{\hat{\bx}^{k}=\mathrm{\mathop {argmin}\limits_{\bx}}  \bF(\bx)-(2\hat{\bLambda}_{\rho\bf I^{*}_{\cG}}^{k}-(\hat{\bV}^{k})^{T})^{T}\bx+ \frac{\rho }{2}\|\bx\|_2^2,}  \\
{\hat{\bLambda}_{\rho\bF^{*}}^{k}=2\hat{\bLambda}_{\rho\bf I^{*}_{\cG}}^{k}-(\hat{\bV}^{k})^{T}-\rho\hat{\bx}^{k}} .
\end{array}}\right.
\end{eqnarray}

According to the updated presented in (\ref{async2}), (\ref{z_k}), and (\ref{x_k}), we can see that updates (\ref{v_k+1}) and (\ref{x_k}) are completely decoupled across all BSs, and they can be directly divided into subproblems for the BSs to solve independently. However,  the update procedure in (\ref{z_k}) cannot be directly performed in a distributed and asynchronized manner. We consider deriving the analytical solution of (\ref{z_k}) via reformulating it  as the following convex optimization problem
\vspace{-0.4cm}
\begin{subequations}
  \begin{align}
    \mathrm{\mathop {argmin}\limits_{\bz}} \quad  &\bz^{T}\hat{\bV^{k}}  + \frac{\rho }{2}\|\bz\|_2^2 \\
    \mbox{ s.t. } \quad \quad   & A\bz  \le \gamma.
  \end{align}
  \label{Zsk_subproblem}
\end{subequations}
\vspace{-0.3cm}
We can get the closed-form solution of (\ref{Zsk_subproblem}) via the (Karush-Kuhn-Tucker) KKT conditions \cite{boyd2004convex} as follows:
\vspace{-0.8cm}
\begin{align}
\hat{\bz}^{k} = [\frac{1 }{\rho}A^{T}(AA^{T})^{-1}A-\frac{1 }{\rho}I_A]\hat{\bV^{k}}+\gamma A^{T}(AA^{T})^{-1}.
\end{align}
In this way, we can easily divide $\frac{1 }{\rho}A^{T}(AA^{T})^{-1}A-\frac{1 }{\rho}I_A$ and $\gamma A^{T}(AA^{T})^{-1}$ into $S$ blocks, so $\hat{\bz}_{s_k}^{k}$-subproblem can be solved in a distributed and asynchronized manner as follows:
\begin{align}
\hat{\bz}_{s_k}^{k} = [\frac{1 }{\rho}A^{T}(AA^{T})^{-1}A-\frac{1 }{\rho}I_A]_{s_k}\hat{\bV}_{s_k}^{k}+[\gamma A^{T}(AA^{T})^{-1}]_{s_k},
\end{align}

To summarize, our distributed asynchronous ADMM algorithm can be performed via the following updates step by step. We describe more details in Algorithm \ref{Algorithm 2}.
\begin{subequations}\label{Eq_Async}
  \begin{align}
  &\hat{\bz}_{s_k}^{k} = [\frac{1 }{\rho}A^{T}(AA^{T})^{-1}A-\frac{1 }{\rho}I_A]_{s_k}\hat{\bV}_{s_k}^{k}+[\gamma A^{T}(AA^{T})^{-1}]_{s_k},\label{asynchronous_z}\\
  &{\hat{\bLambda}_{{\rho\bf I^{*}_{\cG}},{s_k}}^{k}=\hat{\bV}_{s_k}^{k}+\rho\hat{\bz}_{s_k}^{k}},\\
  &\hat{\bx}_{s_k}^{k}=\mathrm{\mathop {argmin}\limits_{\bx_{s_k}}}  \bF_{s_k}(\bx_{s_k})-(2\hat{\bLambda}_{\rho\bf I^{*}_{\cG},{s_k}}^{k}-(\hat{\bV}_{s_k}^{k})^{T})^{T}\bx_{s_k}+ \frac{\rho }{2}\|\bx_{s_k}\|_2^2,  \\
  &{\hat{\bLambda}_{\rho\bF^{*},{s_k}}^{k}=2\hat{\bLambda}_{\rho\bf I^{*}_{\cG},{s_k}}^{k}-(\hat{\bV}_{s_k}^{k})^{T}-\rho\hat{\bx}_{s_k}^{k}},\\
  &\bV_{s_k}^{k+1}=\bV_{s_k}^{k}-\alpha^{k}\mathbf{U}_{s_k}(\hat{\bLambda}_{\rho\bf I^{*}_{\cG},{s_k}}^{k}-\hat{\bLambda}_{\rho\bF^{*},{s_k}}^{k})\label{asynchronous_v}
  \end{align}
\end{subequations}

\begin{algorithm}
\footnotesize
  \caption{Distributed Asynchronized ADMM with Partial Variable Splitting (AsyncADMM)}\label{Algorithm 2}
  \begin{itemize}
    \item[] Initialization: Each BS $s_k$ chooses an initial variable $\bV_{s_k}^0$; $k=1$
    \item[] Set the maximum number of iterations as $K>0$
    \item[] {\bf while} $k \leq K$ {\bf do}
          \begin{itemize}
            \item[] 1. Each BS $s_k$ asynchronously do:
                  \begin{itemize}
                    \item[] 1) Each BS read $\hat{\bV}^{k}$ in the global memory, and calculate according to (\ref{asynchronous_z})-(\ref{asynchronous_v});
                    \item[] 2) Each BS report $\bV_{s_k}^{k+1}$ to the F-orchestrator;
                  \end{itemize}
            \item[] 2.When F-orchestrator receive $\bV_{s_k}^{k+1}$ of any BS, it will do:
                  %\begin{itemize}
                  \begin{itemize}
                    \item[] 1) Update $\bV^{k+1}$ according to $\bV_{s_k}^{k+1}$;
                    \item[] 2) F-orchestrator updates $\hat{\bV}^{k+1}$;
                    \item[] 3) {\bf if} {Stopping criteria met}
                    \item[]  \quad\quad break;
                    \item[] \quad {\bf end if}
                    \item[] 4) Sends the sub-vectors $\bz_s^{k+1}$ and $\bLambda_s^{k+1}$ to the corresponding BS $s$;
                  \end{itemize}
            \item[] 3. $k=k+1$
          \end{itemize}
    \item[] {\bf end while}
  \end{itemize}
\end{algorithm}

\begin{theorem}
\label{Theorem_convergence2}
Let $\bV^{*}$ be the set of optimal values of problem (\ref{Eq_Async}), and $\bLambda^{*}$ be the optimal variable of (\ref{Dual_Prob}). Since
$\bF^{*}(\bLambda)$ and $\bf I^{*}_{\cG}$ are convex functions, $\mathcal{M}_{DRS}$ is a nonexpansive operator and the sequence generated by
 AsyncADMM $(\bV^{k})_{k\geq0}$ can converge to $\bV^{*}$ with probability of 1.
%\begin{itemize}
%  \item[] 1) $\bF^{*}(\bLambda)$ and $\bf I^{*}_{\cG}$ are convex functions, and $\mathcal{T}$ is nonexpansive;
%  \item[] 2) the latency of each update has a positive upper bound;
%  \item[] 3) $\eta_k\in[\eta_{min},\eta_{max}]$, for $\forall \quad 0<\eta_{max}<{{S p_{min}}\over {2\tau \sqrt{k p_{min}}+k}}, and \quad 0<\eta_{min}<\eta_{max}$;
%\end{itemize}

\end{theorem}
\begin{IEEEproof}
Please refer to Appendix C.
\end{IEEEproof}

\section{Case Study: Network Slicing in Smart Transportation}
\label{Section_SIMULATION}
\subsection{Simulation Setup}
\begin{figure}
  \centering
  \includegraphics[width=13cm]{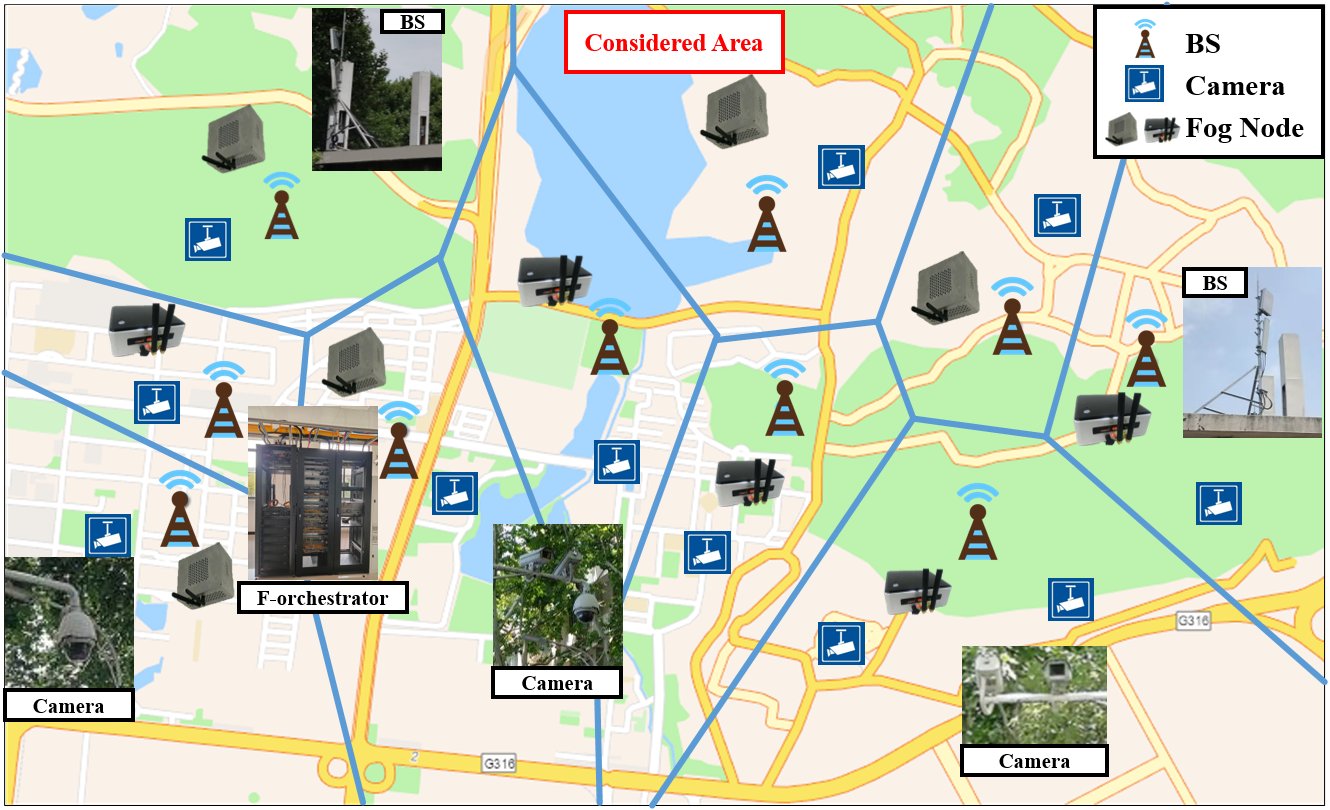}
  \caption{Topology of 5G BSs, control plane, and fog nodes in the considered campus smart-transportation system. }\label{fig2a}
  \vspace{-0.5cm}
\end{figure}

In this section, we evaluate the performance of our proposed distributed network slicing  framework by simulating a smart transportation system inside an university campus supported by a 5G network with fog infrastructure. The 5G network consists of 10 BSs located throughout a 7000-acre university campus as shown in Figure \ref{fig2a}. We assume each BS has been installed with a fog node. The 5G network is connected with a  roadside surveillance system with over 200 cameras located at the main driving route throughout the campus. One of the main challenges of the university transportation system is that many students tend to ignore the guidance of the roadside facility such as traffic light signals and pedestrian signs. It is therefore important to allocate sufficient networking resources and virtualized functions to the smart transportation system so high-precision driving-assistance and pedestrian guidance services can be offered in real time. In this case study, we analyze two-month traffic video collected inside the main roads throughout the campus and use the streaming data rates as well as the portion of the human activity and the moving vehicles in each video frame to infer the spatial and temporal service data rates of two types of services: human-related  (H) service and vehicular-related (V) services. We assume the size of each task related to V and H services are different size  and the task arrival rate at each fog node is given by the service data rate divided by the size of the task associated with each service. We assume the V-service is corresponding to the 5G C-V2X-based roadside assistant related service (e.g., cross-traffic warning and driving assistance service) which according to 5GAA takes around 200-1000 bytes per task \cite{V2X5GWhitePaper}. We also set the task size of H-service as 300 bytes per task. Motivated by the fact that the behavior of the vehicles needs to be collected and analyzed within a shorter time compared to the human activity, we assume the maximum latency that can be tolerated by the V-service is much smaller than that of the H-service, i.e., we set the maximum latency for V and H services as 100 ms  and 500 ms, respectively. It is therefore important for the 5G network to adopt different number of NFs for these two types of services.

\begin{figure}
 \subfigure[Considered Area]{
\begin{minipage}[t]{0.5\linewidth}
\centering
\includegraphics[width=8cm]{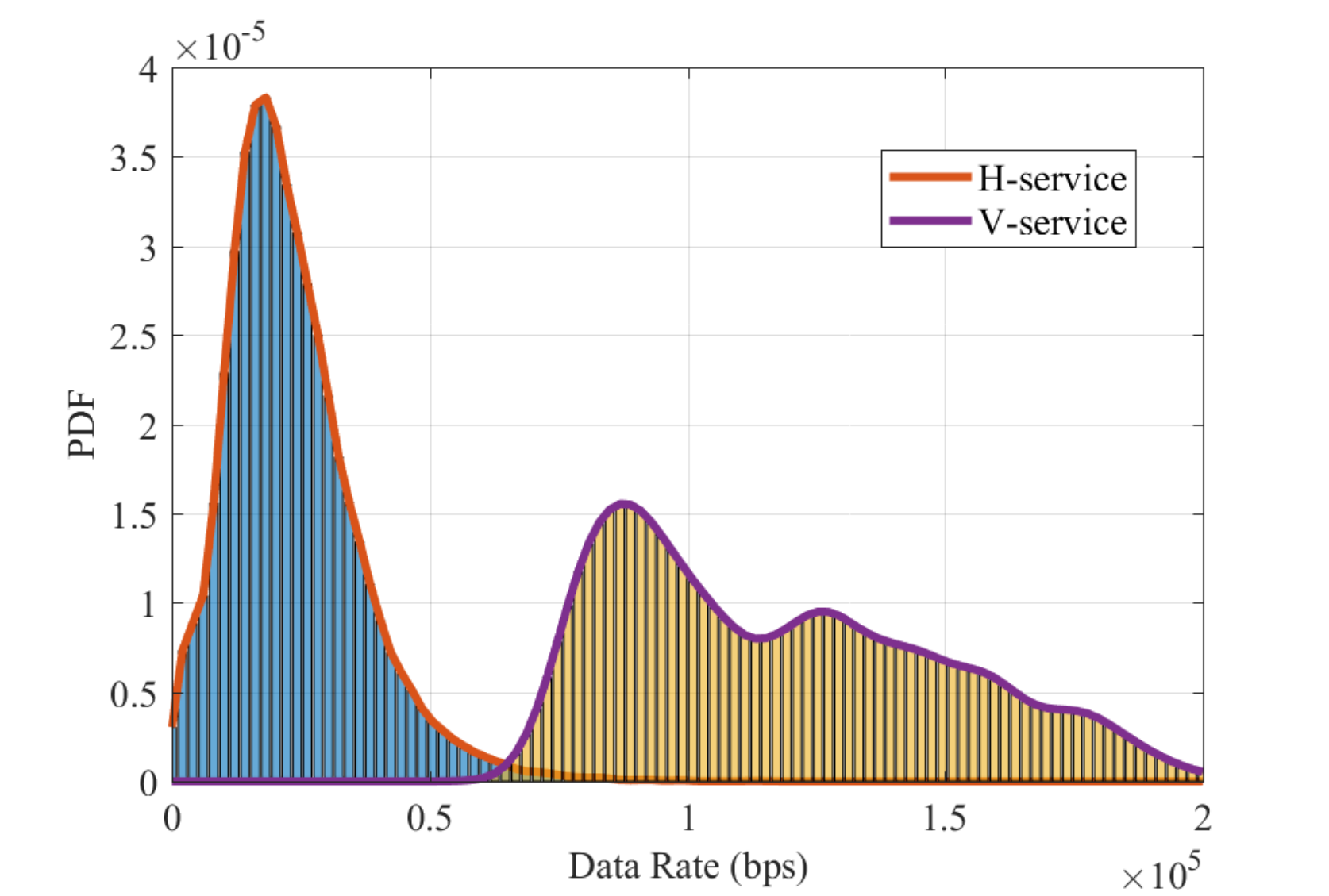}
%\vspace{-0.3in}
%\caption{PDF of H and V Service in area 9}
\label{fig2b}
\end{minipage}
}
\subfigure[All Areas]{
\begin{minipage}[t]{0.5\linewidth}
\centering
\includegraphics[width=8cm]{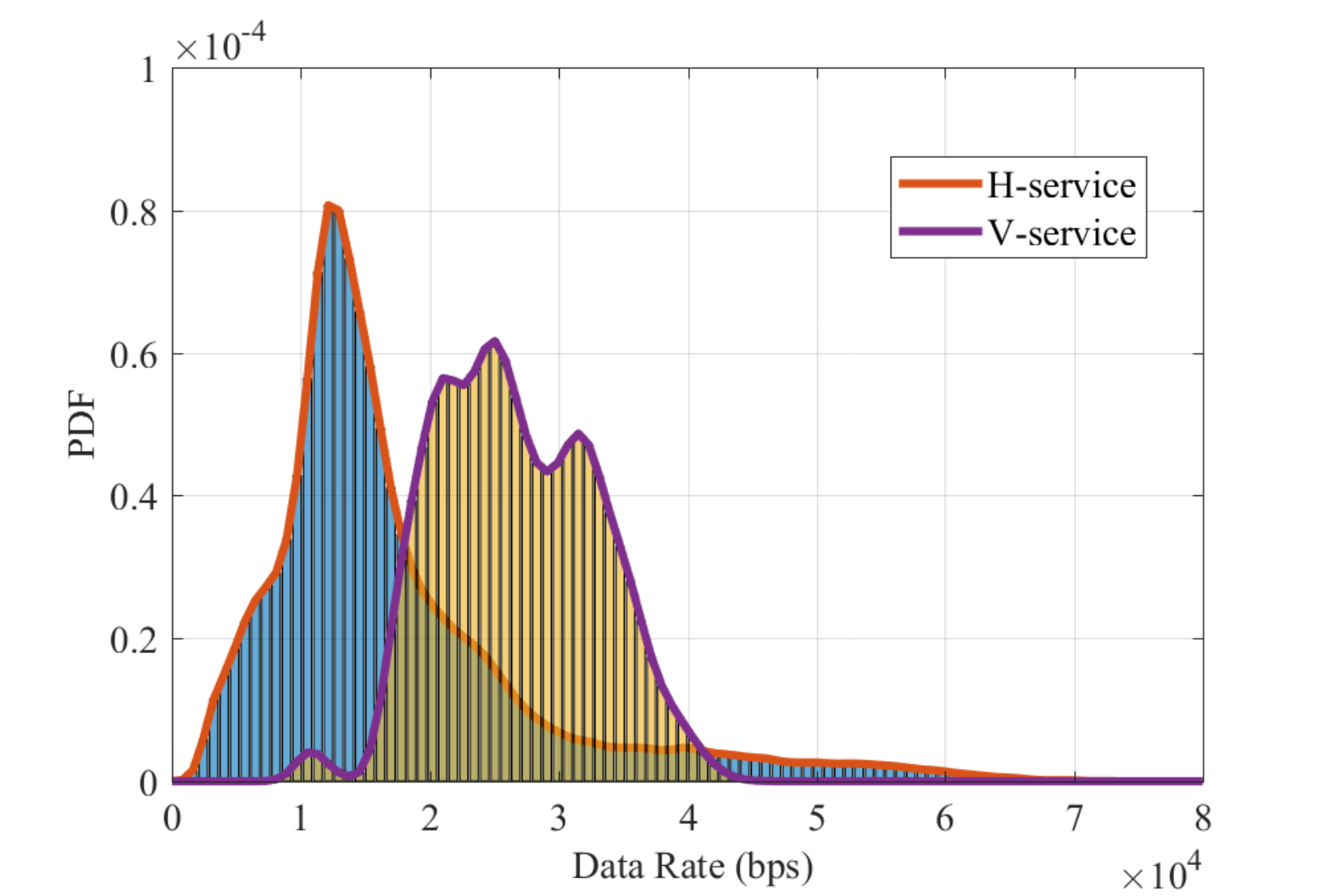}
%\vspace{-0.3in}
%\caption{PDF of H and V Service of all areas}
\label{fig2b1}
\end{minipage}
}
\caption{Empirical PDFs of H and V services}
\end{figure}

The service data rates exhibit significant temporal and spatial variation. In particular, we generate empirical PDFs of both types of services in each BS coverage area using two-month traffic surveillance video collected in all the road-side cameras. The empirical PDF of service data rate in service data rate in one BS coverage area (highlighted in Figure~\ref{fig2b}) as well as PDF of the combined traffic data in the entire campus are presented in Figure~\ref{fig2b1}. We can observe that V-service usually generates a much higher service data rate than the H-service. Also the STD of H-service is always smaller than that of V-service.

To evaluate the temporal variation of the service data rate, we present the mean and STD of the data rates for both V and H services during an 8 hour duration from 9:00 am to 5:00 pm in Figure~8. We can observe that although the data rate of both services can vary at different time and location, they mostly follow very similar peak hours with relatively higher data traffic from 9:30 am to 12:30 am and also non-peak hours with low data rate from 12:30am to 3:00pm. In the rest of this paper, we evaluate the network slicing performance using the service data rate for both peak and non-peak hours.

%\begin{figure}
 %\subfigure[]{
%\begin{minipage}[t]{0.5\linewidth}
%\centering
%\includegraphics[width=4.5cm]{area_final.pdf}
%\vspace{-0.3in}
%\label{fig2a}
%\end{minipage}
%}
%\subfigure[]{
%\begin{minipage}[t]{0.5\linewidth}
%\centering
%\includegraphics[width=4.5cm]{area1.pdf}
%%\vspace{-0.3in}
%
%\label{fig2a1}
%\end{minipage}
%}
%\caption{Distribution of BSs}
%\end{figure}

%\begin{figure}
%\centering
%\includegraphics[width=7cm]{PDF.pdf}
%%\vspace{-0.3in}
%\caption{PDF of Frame Size}
%\label{fig2b}
%
%\vspace{-0.3in}
%\end{figure}
%
%\begin{figure}
%\begin{minipage}[t]{0.5\linewidth}
%\centering
%\includegraphics[width=7cm]{V_SERVICE.pdf}
%%\vspace{-0.3in}
%\caption{Mean and Standard Deviation of Data Rate of V Service}
%\label{fig_carwindow}
%\end{minipage}
%%
%\begin{minipage}[t]{0.5\linewidth}
%
%\centering
%\includegraphics[width=7cm]{H_SERVICE.pdf}
%%\vspace{-0.3in}
%\caption{Mean and Standard Deviation of Data Rate of H Service}
%\label{fig_humanwindow}
%\end{minipage}
%
%\vspace{-0.3in}
%\end{figure}
\begin{figure}
 \subfigure[Service V]{
\begin{minipage}[t]{0.5\linewidth}
\centering
\includegraphics[width=8cm]{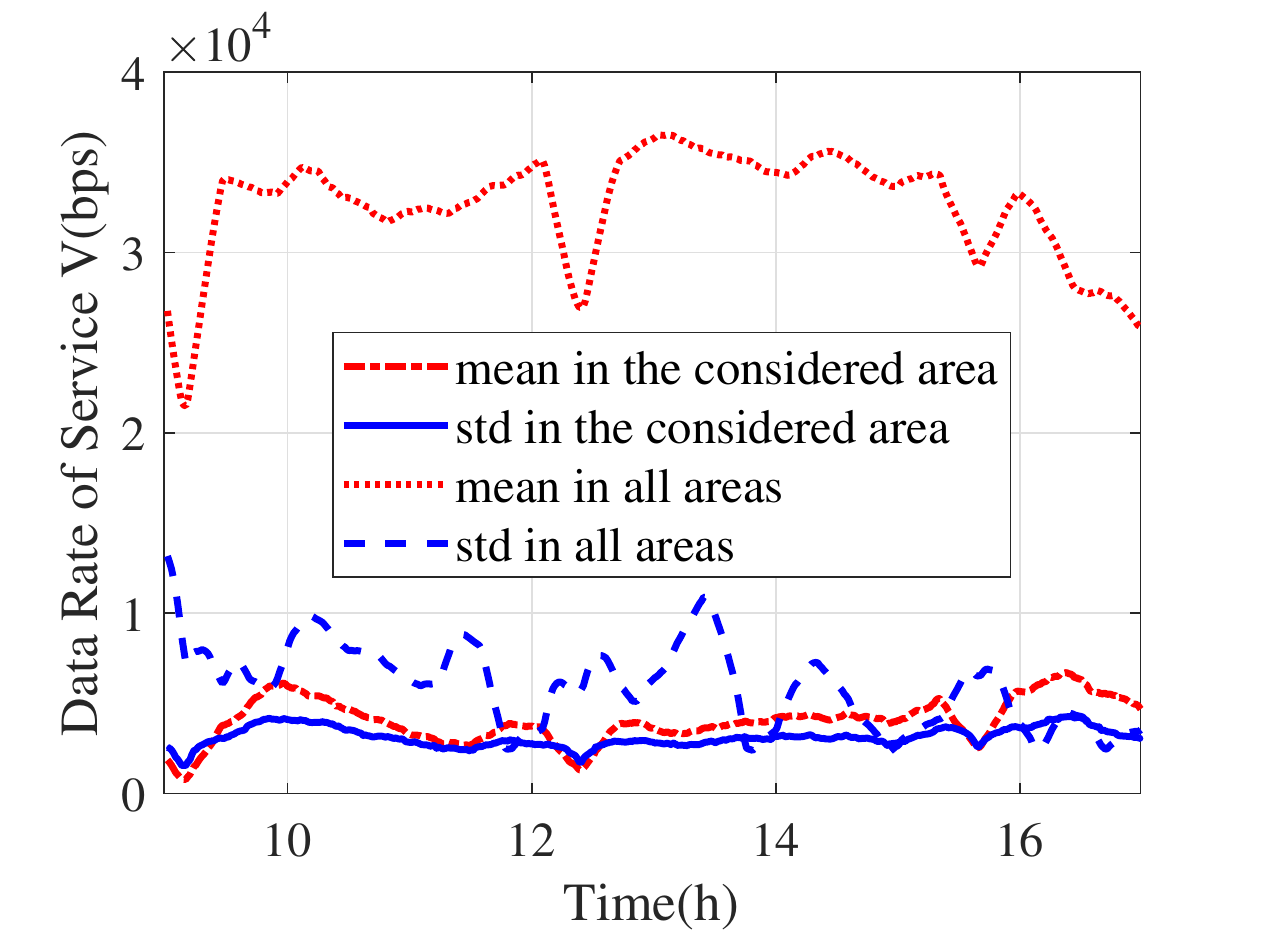}
%\vspace{-0.3in}
%\caption{PDF of H and V Service in area 9}
\label{fig_carwindow}
\end{minipage}
}
\subfigure[Service H]{
\begin{minipage}[t]{0.5\linewidth}
\centering
\includegraphics[width=8cm]{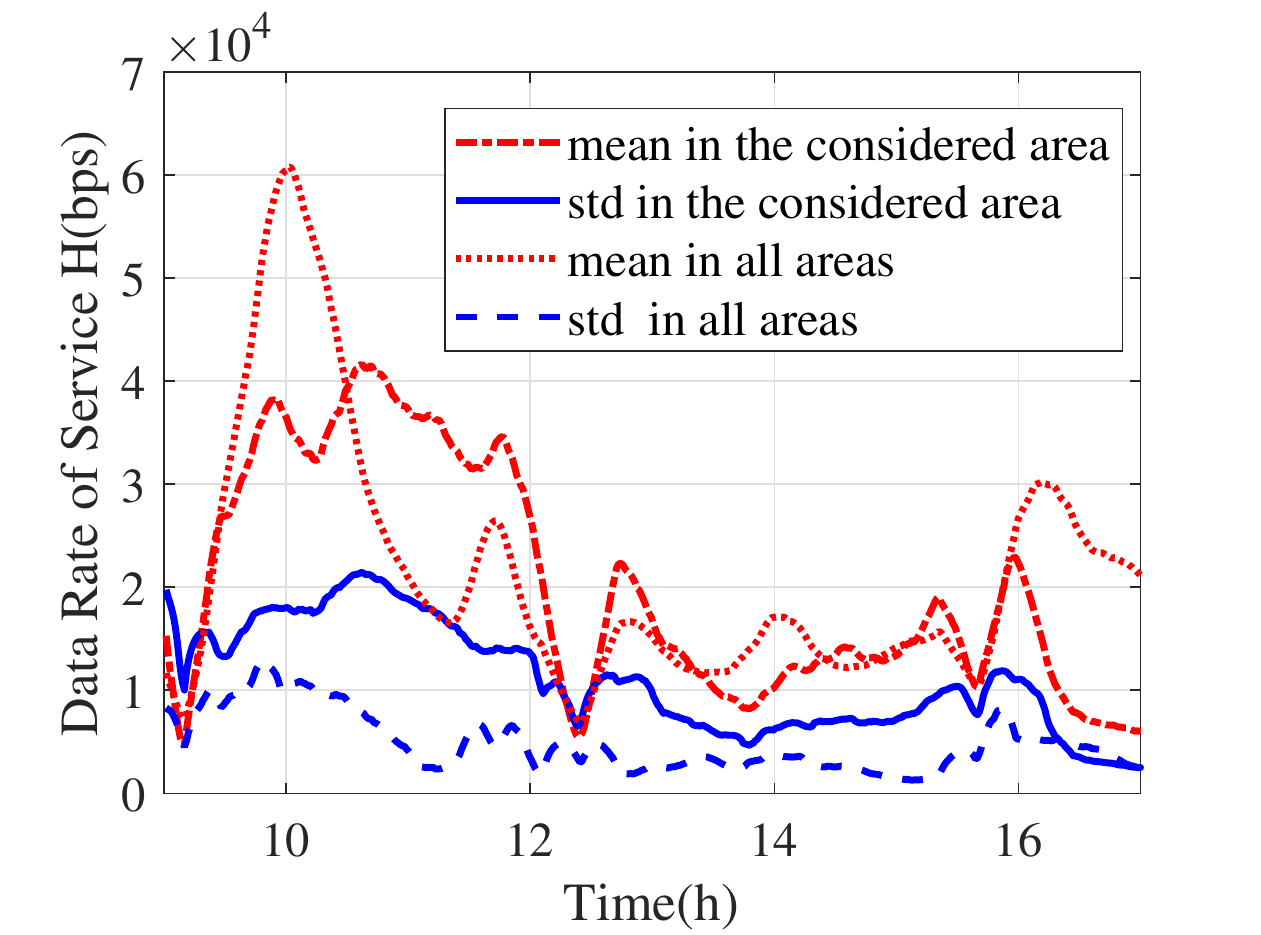}
%\vspace{-0.3in}
%\caption{PDF of H and V Service of all areas}
\label{fig_humanwindow}
\end{minipage}
}
\caption{Mean and STD of the data rates of both V and H services}
\end{figure}
%\begin{figure}
%\begin{minipage}[t]{0.5\linewidth}
%\centering
%\includegraphics[width=4.5cm]{1_Pren.pdf}
%\vspace{-0.3in}
%\caption{}
%\label{fig_Pren}
%\end{minipage}
%%
%\begin{minipage}[t]{0.45\linewidth}
%
%\centering
%\includegraphics[width=4cm]{1_Iren.pdf}
%\vspace{-0.3in}
%\caption{}
%\label{fig_Iren}
%\end{minipage}
%
%%\vspace{-0.3in}
%\end{figure}
%
%\begin{figure}
%\begin{minipage}[t]{0.5\linewidth}
%\centering
%\includegraphics[width=4.5cm]{1_Pche.pdf}
%\vspace{-0.3in}
%\caption{}
%\label{fig_Pche}
%\end{minipage}
%%
%\begin{minipage}[t]{0.45\linewidth}
%
%\centering
%\includegraphics[width=4cm]{1_Iche.pdf}
%\vspace{-0.3in}
%\caption{}
%\label{fig_Iche}
%\end{minipage}
%
%%\vspace{-0.3in}
%\end{figure}
%
%\begin{figure}
%\begin{minipage}[t]{0.5\linewidth}
%\centering
%\includegraphics[width=4.5cm]{1_Poth.pdf}
%\vspace{-0.3in}
%\caption{}
%\label{fig_Poth}
%\end{minipage}
%%
%\begin{minipage}[t]{0.45\linewidth}
%
%\centering
%\includegraphics[width=4cm]{1_Ioth.pdf}
%\vspace{-0.3in}
%\caption{}
%\label{fig_Ioth}
%\end{minipage}
%
%%\vspace{-0.3in}
%\end{figure}
\vspace{-0.5cm}
\subsection{Simulation Results}
\begin{comment}
\begin{table}
\caption{List of Notations}
\centering
\begin{tabular}{c|p{5.8 cm}}
\hline
Symbol & Definition \\
\hline
  $N$ & 3 \\
  $S$ & 2 \\
  $\beta_s(M)$ & 40 \\
  $b_0(kb)$ & 25 \\
  $\gamma$ & 500\\
  $d_1,d_2,d_3(kb)$ & 5,32,100\\
  $\lambda_{1,1},\lambda_{1,2},\lambda_{1,3}$ &6,10,14\\
  $\lambda_{2,1},\lambda_{2,2},\lambda_{2,3}$ &18,22,26\\
  $\overline{Tp_{l}},\overline{Tp_{2}},\overline{Tp_{3}}(s)$ &0.5,0.1,0.05\\
\hline
\label{table 2}
\end{tabular}
\end{table}
\end{comment}

%\begin{figure}
%\begin{minipage}[t]{0.4\linewidth}
%\centering
%\vspace{-0.25in}
%\includegraphics[width=4.3cm]{area2.pdf}
%\vspace{-0.25in}
%\caption{Distribution of BSs}
%\label{fig2a}
%\end{minipage}%
%
%\begin{minipage}[t]{0.4\linewidth}
%\centering
%\vspace{0.06in}
%\includegraphics[width=4cm]{table1.pdf}
%\vspace{-0.27in}
%\caption{Average Coverage Area of BSs}
%\label{fig2b}
%\end{minipage}%
%\vspace{-0.23in}
%
%\end{figure}
In this section, we mainly present our simulation results. We evaluate the performance of our proposed algorithms, and compare the joint slicing involving both bandwidth of BSs and processing powers of fog nodes with two other network slicing scenarios: bandwidth slicing and computational resource slicing. The major parameters and their values are presented in Table~\ref{Table_Para}.

\begin{table}[!t]
  \centering
    \caption{Simulation Parameters}
  \begin{tabular}{|p{5.5cm}|p{2cm}||p{5cm}|p{2.8cm}|}
    \hline
       Parameter      & Value             & Parameter         & Value                \\
    \hline \hline
    Bandwidth of  BS         & $60$ (MHz)           & Processing power of fog node     & $10000$ (tasks/s)                                         \\
    \hline
   Task size of H-service ($S^{H}$)     & $300$ (bytes)      & Task size of V-service ($S^{V}$)     & $500$ (bytes)                            \\
    \hline
   Transmission latency threshold     &   $0.5$ (s)    & Queuing delay threshold     & $0.1$ (s)\\
    \hline
  \end{tabular}

  \label{Table_Para}
\end{table}

%\begin{figure*}[!htbp]
%\setcounter{figure}{7}
%\begin{figure}
% \centering
% \includegraphics[width=7cm]{fig_Iteration.pdf}
%% \vspace{-0.3in}
% \caption{Comparison of Interior-Point and Algorithm 1}
% \label{fig3}
%\end{figure}%
%\begin{figure}
%\centering
%\includegraphics[width=8cm]{fig_Iterationnew.pdf}
%\label{fig3}
%\caption{Comparison of Interior-Point, Algorithm 1 and Algorithm 2}
%\end{figure}
\begin{figure}
 \centering
 \includegraphics[width=10cm]{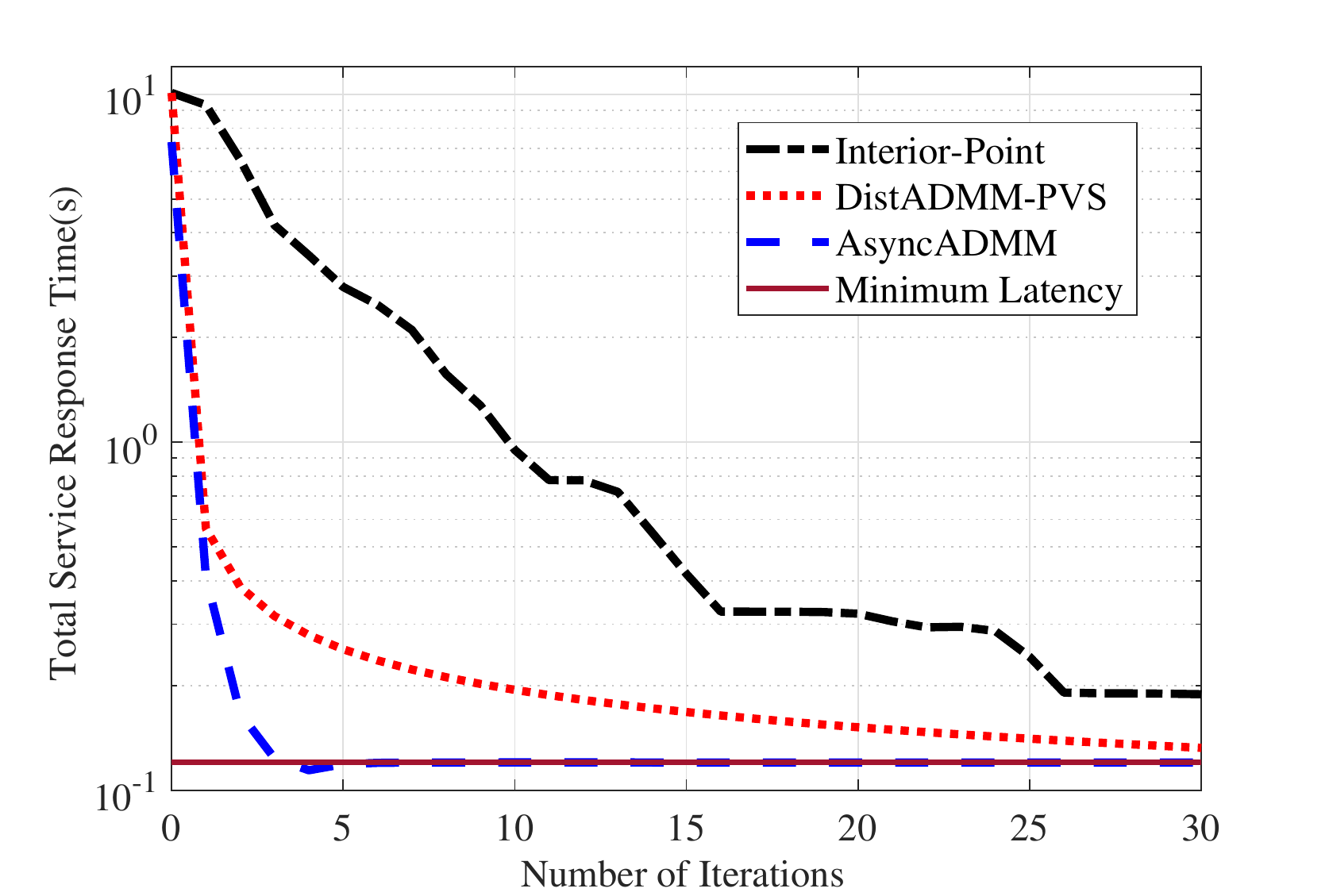}
% \vspace{-0.3in}
 \caption{Convergence performances of proposed algorithms. }
 \label{fig3}
 \vspace{-1.5cm}
\end{figure}
\begin{figure}
 \subfigure[Peak Hour]{
\begin{minipage}[t]{0.5\linewidth}
\centering
\includegraphics[width=8cm]{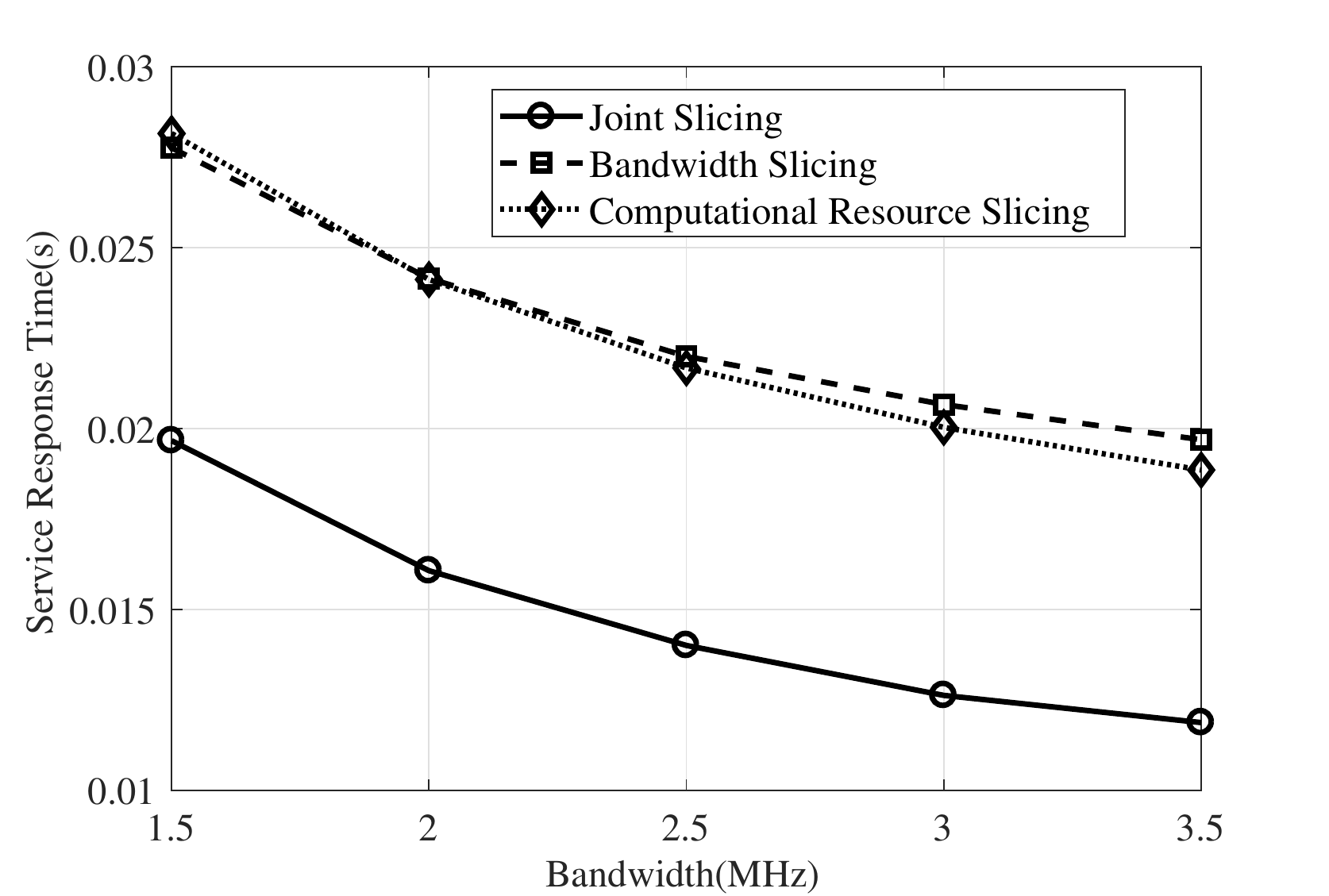}
%\vspace{-0.3in}
%\caption{PDF of H and V Service in area 9}
\label{fig4a}
\end{minipage}
}
\subfigure[Non-peak Hour]{
\begin{minipage}[t]{0.5\linewidth}
\centering
\includegraphics[width=8cm]{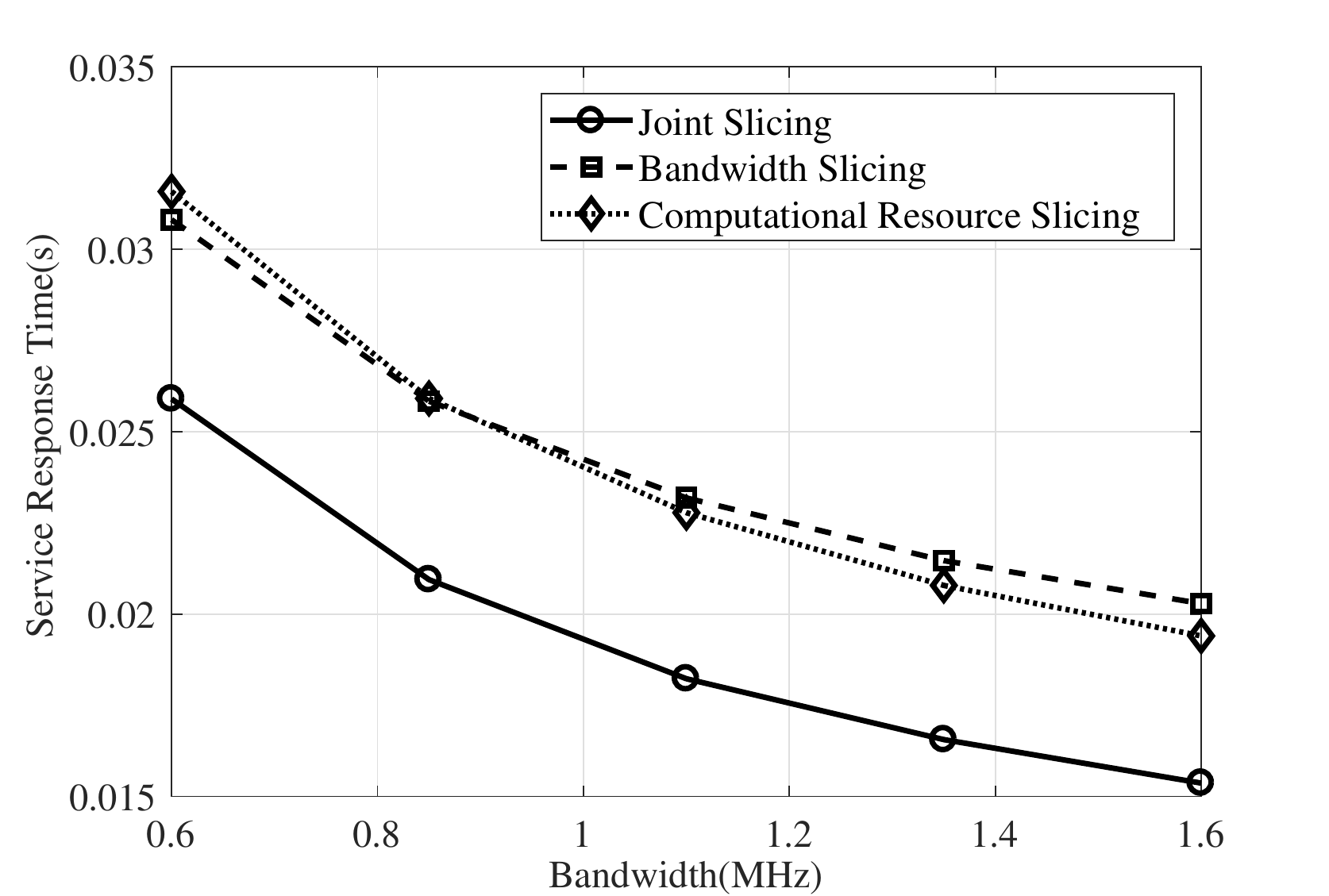}
%\vspace{-0.3in}
%\caption{PDF of H and V Service of all areas}
\label{fig4b}
\end{minipage}
}
\caption{Comparison of the three slicing frameworks under different bandwidths of BSs}
\end{figure}
%\begin{figure}
% \centering
% \includegraphics[width=7cm]{fig_COM.pdf}
%% \vspace{-0.3in}
% \caption{Comparison of Three Architectures under Different Processing Rate of Each Fog Node}
% \label{fig5}
%\end{figure}%
\begin{figure}
 \subfigure[Peak Hour]{
\begin{minipage}[t]{0.5\linewidth}
\centering
\includegraphics[width=8cm]{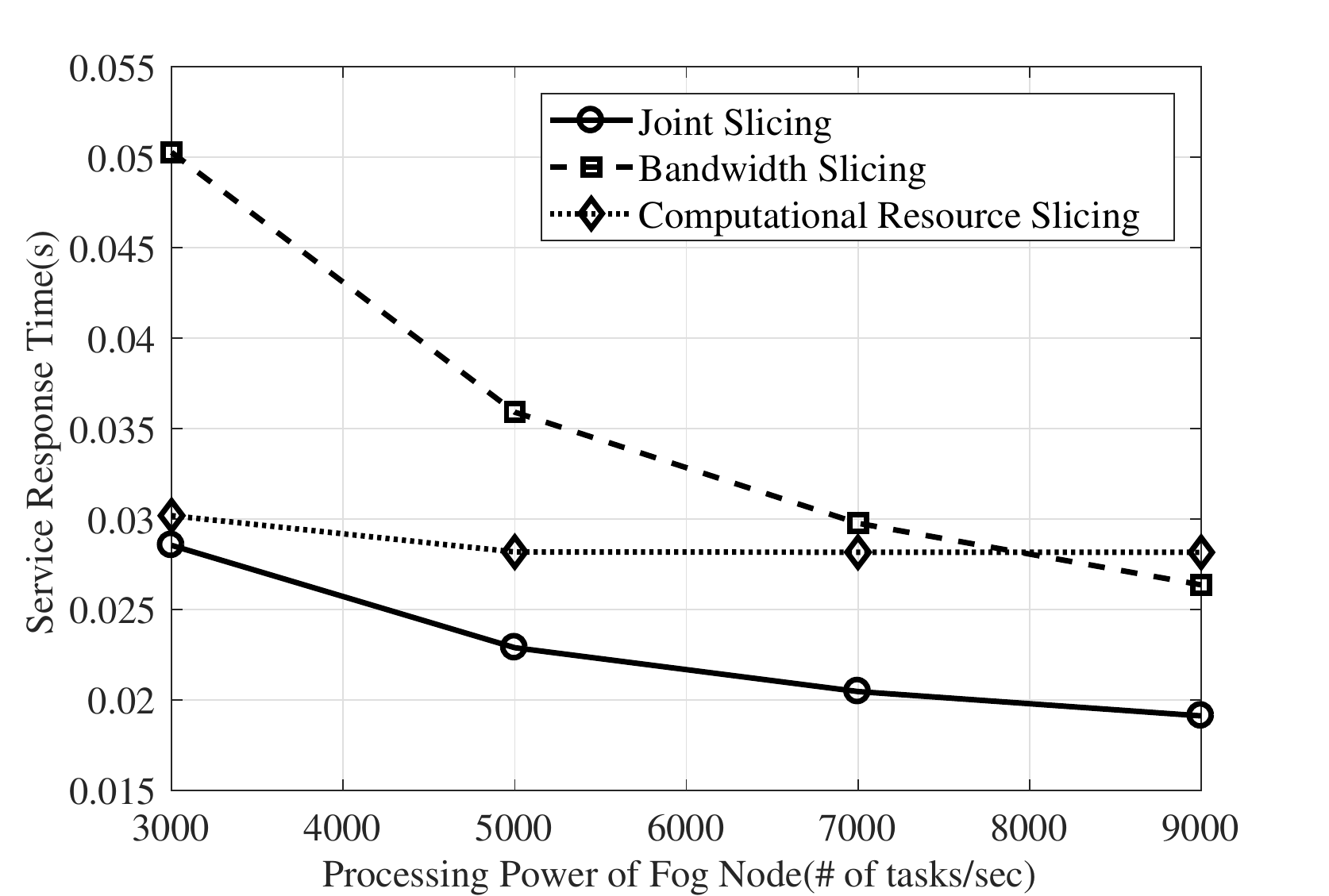}
%\vspace{-0.3in}
%\caption{PDF of H and V Service in area 9}
\label{fig5a}
\end{minipage}
}
\subfigure[Non-peak Hour]{
\begin{minipage}[t]{0.5\linewidth}
\centering
\includegraphics[width=8cm]{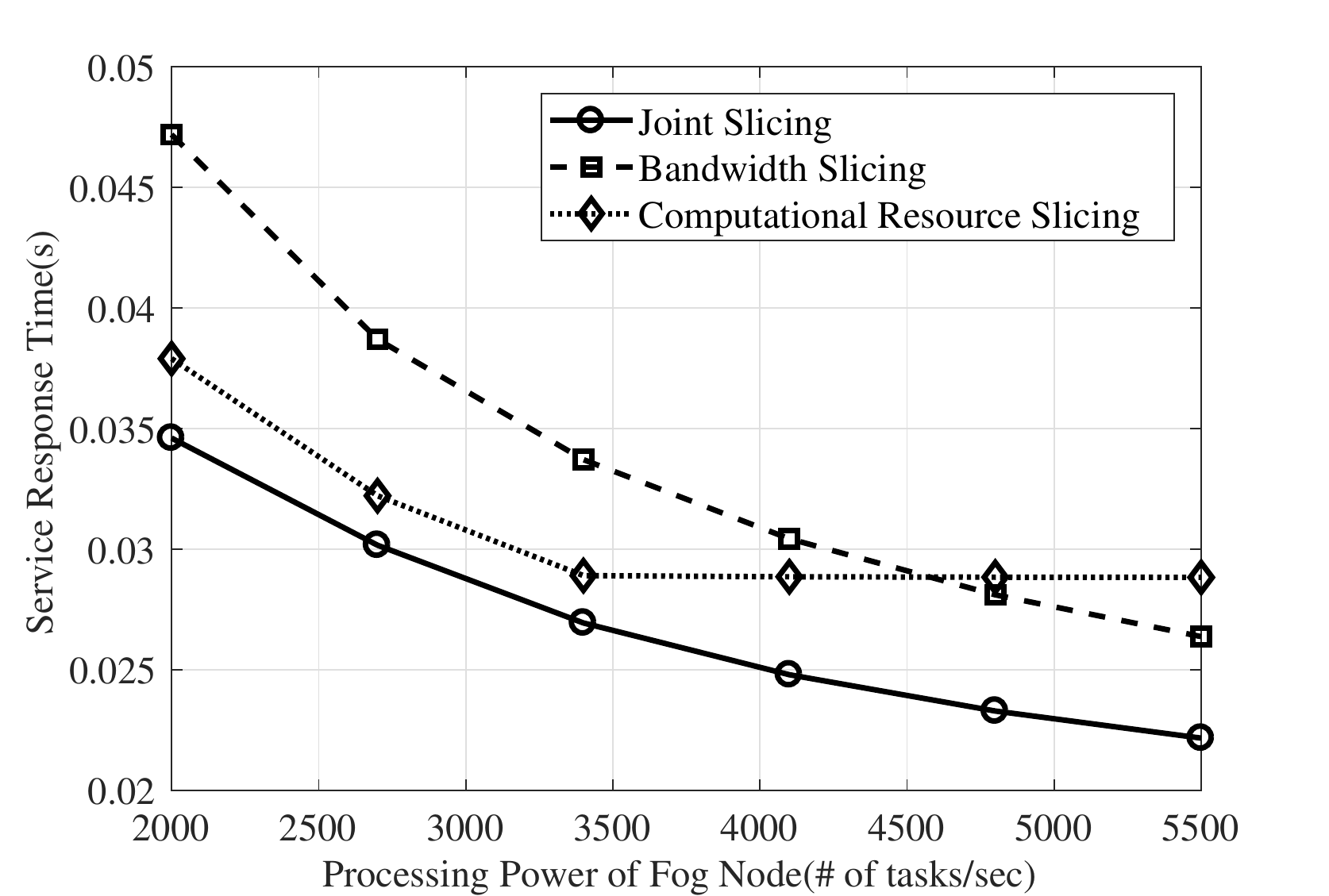}
%\vspace{-0.3in}
%\caption{PDF of H and V Service of all areas}
\label{fig5b}
\end{minipage}
}
\caption{Comparison of the three slicing frameworks under different processing rates of each fog node}
\end{figure}

%\begin{figure}
% \centering
% \includegraphics[width=7cm]{fig_theta.pdf}
%% \vspace{-0.3in}
% \caption{Comparison of Three Architectures under Different $\theta$}
% \label{fig6}
% \vspace{-0.33in}
%\end{figure}
\begin{figure}
 \subfigure[Peak Hour]{
\begin{minipage}[t]{0.5\linewidth}
\centering
\includegraphics[width=8cm]{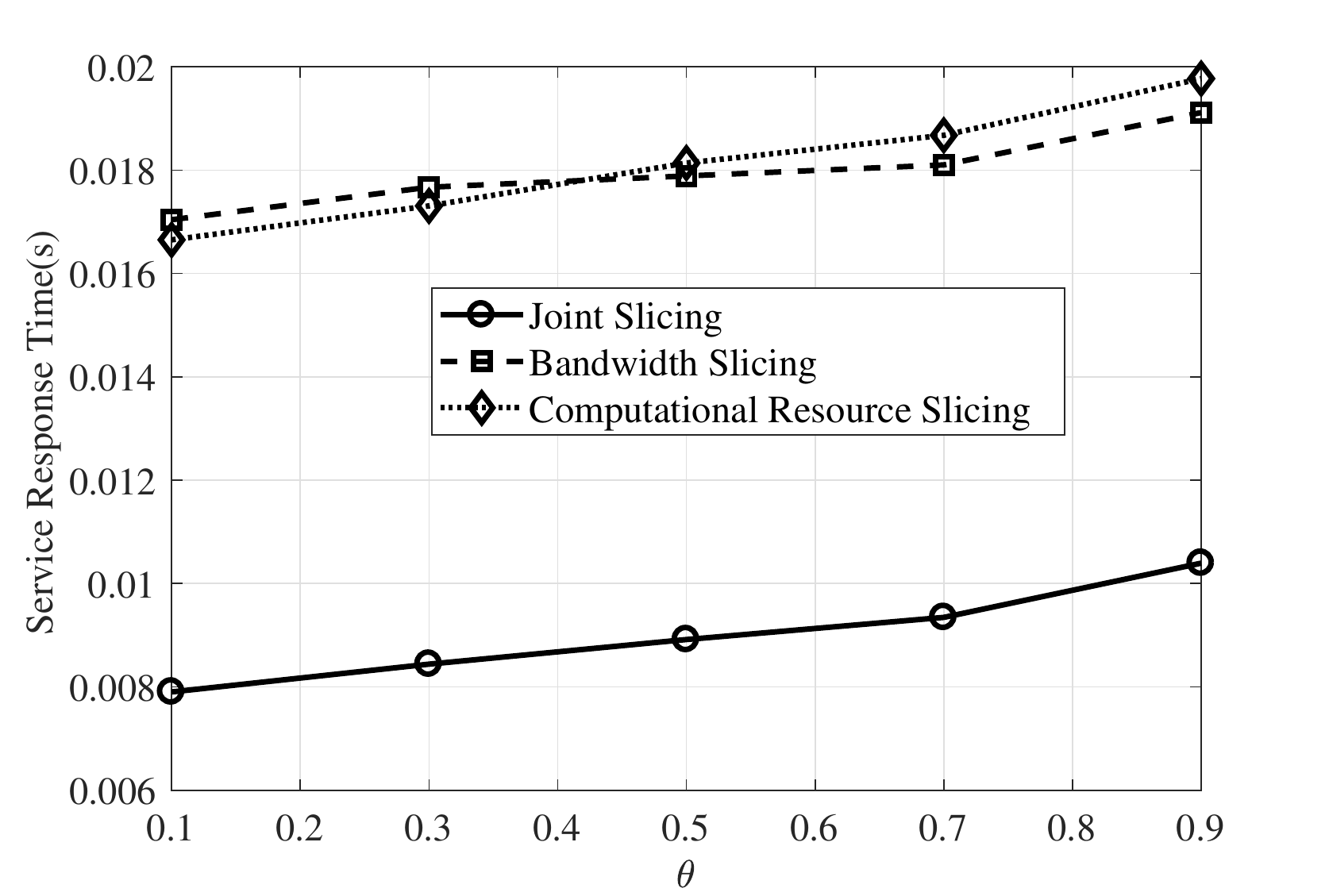}
%\vspace{-0.3in}
%\caption{PDF of H and V Service in area 9}
\label{fig6a}
\end{minipage}
}
\subfigure[Non-peak Hour]{
\begin{minipage}[t]{0.5\linewidth}
\centering
\includegraphics[width=8cm]{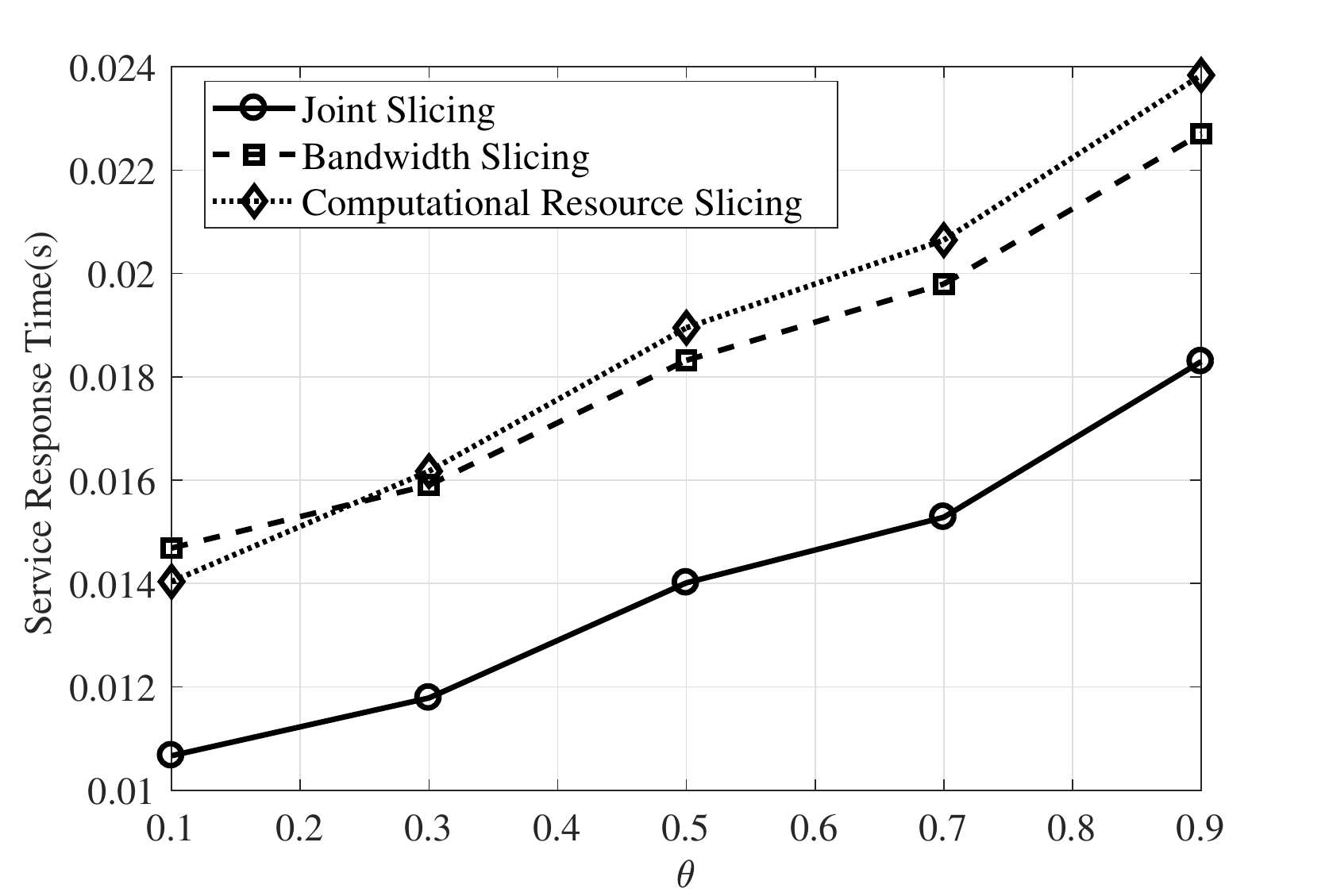}
%\vspace{-0.3in}
%\caption{PDF of H and V Service of all areas}
\label{fig6b}
\end{minipage}
}
\caption{Comparison of  the three slicing frameworks  under different confidence levels $\theta$}
\end{figure}
%\begin{figure}
%\begin{spacing}{-0.6}
%%%行间距变为single-space
%\end{spacing}
%\centering
%\vspace{-0.02in}
%\includegraphics[width=4cm]{fig_Iteration.pdf}
%\vspace{-0.12in}
%\caption{Comparison of Interior-Point and Algorithm 1} % and random pairing.}
%\label{fig3}
%
%\end{figure}

We first evaluate the convergence performance of Algorithm \ref{Algorithm 1} (DistADMM-PVS) and Algorithm \ref{Algorithm 2} (AsyncADMM). In Figure \ref{fig3}, we compare the interior-point algorithm with our proposed Algorithm \ref{Algorithm 1} and Algorithm \ref{Algorithm 2} under different number of iterations. The interior-point algorithm has been widely applied in communication network systems \cite{8382242}. We can observe that our proposed Algorithm \ref{Algorithm 1} and Algorithm \ref{Algorithm 2} can converge to a close neighborhood of the minimum latency within the first few iterations, and our proposed algorithms can offer much faster convergence performance compared to the interior-point algorithm. We can also observe that Algorithm \ref{Algorithm 2} can converge to the optimal value faster than Algorithm \ref{Algorithm 1}, which shows that Algorithm \ref{Algorithm 2} can effectively reduce the waiting time before each iteration of RO that occurs in Algorithm \ref{Algorithm 1}.

%\begin{figure}
%\begin{minipage}[t]{0.5\linewidth}
%\centering
%\vspace{-0.15in}
%\includegraphics[width=4cm]{area_comparison.pdf}
%\vspace{-0.1in}
%\caption{Comparison of Three Architectures in Different Areas}
%\label{fig_area_comparison}
%\end{minipage}
%%
%\begin{minipage}[t]{0.45\linewidth}
%
%\centering
%\vspace{-0.15in}
%\includegraphics[width=4cm]{fig_BW1.pdf}
%\vspace{-0.3in}
%\caption{Comparison of Three Architectures under Different Bandwidth of Each BS of Area1}
%\label{fig4}
%\end{minipage}
%\end{figure}
%
%\begin{figure}
%\begin{minipage}[t]{0.5\linewidth}
%\centering
%\includegraphics[width=4cm]{fig_COM1.pdf}
%\vspace{-0.17in}
%\caption{Comparison of Three Architectures under Different Processing Rate of Each Fog Node of Area1}
%\label{fig5}
%\end{minipage}
%%
%\begin{minipage}[t]{0.45\linewidth}
%
%\centering
%\includegraphics[width=4cm]{fig_theta1.pdf}
%\vspace{-0.33in}
%\caption{Comparison of Three Architectures under Different $\theta$ of Area1}
%\label{fig6}
%\end{minipage}
%
%\vspace{-0.32in}
%\end{figure}

%We compare three architectures in different kinds of areas, as shown in Figure \ref{fig_area_comparison}. We can observe that our proposed joint slicing architecture performs better than other network slicing architecture in all three kinds of areas, which proves that our architecture has significant regional applicability, and can apply to different areas. We have the same characteristics for three architectures in all areas, so we only cover simulation results of area 1 in detail, as shown in Figure \ref{fig4}, Figure \ref{fig5} and Figure \ref{fig6}.

We compare the joint slicing, bandwidth slicing and computational resource slicing architectures in peak hour and non-peak hour. We can observe that our proposed joint slicing architecture performs better than other network slicing architecture in peak and non-peak data rate of H-service and V-service, which proves that our architecture has significant applicability, and can apply to different types of data rate conditions. We have the same characteristics for three architectures in peak and non-peak hour, so we uniformly introduce simulation results of peak and non-peak hour in detail.

In Figure \ref{fig4a} and \ref{fig4b}, we fix the processing power reserved for each fog node and the value of $\theta$ to evaluate the service response time under different bandwidth reserved for each BS. We can observe that the service response time decreases with the total reserved bandwidth. We can also observe that when the bandwidth of BSs is limited, the bandwidth slicing offers a better performance than computational resource slicing. However, as the bandwidth of each BS increases, the service response time of computational resource slicing starts to decrease much faster than that of the bandwidth slicing. This is because in our simulation, we fix the computational resource. In this case, when each BS has limited bandwidth, the communication latency dominates the overall latency. Therefore, applying bandwidth slicing to reduce the communication latency can have a higher impact than optimizing the computational resources in fog nodes for reducing the overall service response time. When the bandwidth of each BS becomes sufficient, the queuing delay will dominate overall latency. In this case, the computational resource slicing will become more useful to reduce the service response time.

In Figure \ref{fig5a} and \ref{fig5b}, we fix the bandwidth reserved for each BS and value of $\theta$ to compare the service response time of three network slicing scenarios under different processing power reserved for each fog node. We can observe that the service response time decreases with the processing power reserved for each fog node. Similarly, we observe that when the computational resource of fog nodes is limited, the computational resource slicing offers a better performance than bandwidth slicing. However, as the computational resource of each fog node becomes sufficient, bandwidth slicing starts to decrease faster than the computational resource slicing. This is because in Figure \ref{fig5a} and \ref{fig5b}, the bandwidth has been fixed. When each fog node has limited processing power, the queuing delay dominates the overall latency. When the processing power of each fog node increases, the communication latency starts to dominate the overall latency.

Similarly, in Figure \ref{fig6a} and \ref{fig6b}, we fix the processing power reserved for each fog node and the bandwidth reserved for each BS to compare the service response time under different values of $\theta$. We observe that the service response time increases with $\theta$. This is because when $\theta$ increases, the total number of task units that need to be transported by BSs and calculated by fog nodes becomes larger. This will cause higher communication latency and queuing delay, resulting in a higher service response time. We also observe that as $\theta$ increases, the service response time of bandwidth slicing starts to increase much slower than that of the computational resource slicing. This is because we fix both processing power reserved for each fog node and the bandwidth reserved for each BS. In this case, the number of task units for each service from each BS increases with $\theta$. When $\theta$ is small, bandwidth allocated to each task unit is sufficient and the queuing delay dominates the overall latency. When $\theta$ becomes larger, bandwidth allocated to each task unit is limited and the communication latency will start to dominate the overall latency.
\begin{comment}
\begin{figure}
  \centering
  \includegraphics[width=8cm]{service_bw.pdf}
  \caption{Service Response Time of each slice under different bandwidth of each BS}
  \label{fig6}
\end{figure}

\begin{figure}
  \centering
  \includegraphics[width=8cm]{service_com.pdf}
  \caption{Service Response Time of each slice under different process rate of all fog nodes}
  \label{fig7}
\end{figure}
\end{comment}
\vspace{-1cm}
\section{Conclusion}
\label{Section_Conclusion}
This paper investigated network slicing for IoT networks supported by 5G and fog computing networks. A novel distributed network slicing framework was proposed based on a new control plane entity, F-orchestrator. In this framework, each BS does not have to share any local information such as service request and resource availability with other BSs nor F-orchestrator. Two distributed algorithms based on DistADMM-PVS and AsynADMM were proposed under our proposed framework. We proved that both algorithms can converge to the global optimal solution with the required QoS for all supported services. we consider a smart transportation system supported by a 5G network deployed in a university campus as a case study to evaluate the performance of our proposed framework. Our results show that our proposed distributed network slicing framework offers significant improvement on service latency performance for all supported services.

\appendices
\section{Proof of Theorem \ref{Theorem_Convex}}
\label{Theorem_1}

For the convexity, it can be directly shown that set $\cG_s, \forall s \in \cal S$ and half-space $\cG$ are all convex sets, and their intersection which is the feasible set of problem (\ref{Sep_Prob}) is also a convex set. We can also show that within the feasible set of problem (\ref{Sep_Prob}), the second derivative of $f(\bx_{s })$ is always positive which means that it is convex. Due to the fact that summation preserves convexity, we can then prove that $\cal L_{\rho} (\bx, \bz, \bLambda)$ is convex.

To prove (\ref{Lag}) is partially separable, we rewrite the augmented Lagrangian in (\ref{Lag}) as follows
%\begin{align}
%   \cal L_{\rho} & (\bx_1, ..., \bx_S, \bz, \bLambda)= \sum_{s\in {\cal S}} \left\{ f(\bx_{s }) + {\bf I}_{\cG_s} \left(\bx_{s} \right)  \nonumber \\
%   & +\bLambda_s^T(\bx_s-\bz_s) + \frac{\rho}{2} ||\bx_s-\bz_s||_2^2 \right\}  + \bf I_{\cG} \left(\bz \right),
%\end{align}
\begin{eqnarray}
 {\cal L}_{\rho} (\bx_1, ..., \bx_S, \bz, \bLambda)= \sum_{s\in {\cal S}} \left\{ f(\bx_{s }) + {\bf I}_{\cG_s} \left(\bx_{s} \right) \right.
  +\bLambda_s^T(\bx_s-\bz_s) + \frac{\rho}{2} ||\bx_s-\bz_s||_2^2 \left.\right\}  + \bf I_{\cG} \left(\bz \right),\label{Lp}
 \end{eqnarray}
From (\ref{Lp}), we can observe that ${\cal L}_{\rho} (\bx_1, ..., \bx_S, \bz, \bLambda)$ can be partially separated across $\bx_s$ for $s\in {\cal S}$.
 This concludes the proof.

\section{Proof of Theorem \ref{Theorem_2}}
\label{AppendixA}
Since the distributed sub-problems specified in (\ref{DistADMM_x}) is equivalent to the centralized $\bx$-update in (\ref{CADMM_x}), the convergence property of our proposed DistADMM-PVS algorithm directly follows that of the standard ADMM approach \cite{boyd2011distributed}.

Let $\sum_{s\in {\cal S}} \left\{ f(\bx_{s }) + {\bf I}_{\cG_s} \left(\bx_{s} \right) \right\} =\bf F(\bx)$. Proving the convergence of our algorithm is equivalent to prove the primal residual $r^{k}\rightarrow0$, the dual residual $s^{k}=\rho(z^{k}-z^{k-1})\rightarrow0$, and $p^{k}=\bf F(\bx^{k})+\bf I_{\cG}(\bz^{k})\rightarrow p^{*}$. $\bf F$, $\bf I_{\cG}$ are closed, proper and convex, let $(\bx^{*}, \bz^{*}, \bLambda^{*})$ be the saddle point of $L_0$, and we define a $Lyapunov function$ $I_k$ as follows
\begin{align}
I^{k} = {1\over \rho}\|\bLambda^{k} - \bLambda^{*}\|_2^2+ \rho\|\bz^{k} - \bz^{*}\|_2^2.
\end{align}

The proof has three main inequalities, the first is as follows
\begin{align}
I^{k+1} \leq I^{k}-\rho\|r^{k+1}\|_2^2+ \rho\|\bz^{k} - \bz^{k}\|_2^2,\label{inequaly1}
\end{align}

Because $I^{k} \leq I^{0}$, so we have
\begin{align}
\rho \sum _{k=0}^{\infty}(\|r^{k+1}\|_2^2+ \|\bz^{k+1} - \bz^{k}\|_2^2)\leq I^{0},
\end{align}
which means that if $k\rightarrow\infty$, then $r^{k}\rightarrow0$ and $\bz^{k+1} - \bz^{k}\rightarrow0$, so $\rho(\bz^{k+1} - \bz^{k})\rightarrow0$, that is $s^{k}\rightarrow0$.

The second inequality is as follows
\begin{align}
p^{k+1}-p^{*} \leq -(\bLambda^{k+1})^{T}r^{k+1}-\rho((\bz^{k+1} - \bz^{k})^{T}(-r^{k+1}+\bz^{k+1} - \bz^{*})).
\end{align}

So the third inequality is as follows
\begin{align}
p^{*}-p^{k+1} \leq (\bLambda^{*})^{T}r^{k+1}.
\end{align}
So we have $lim_{k\rightarrow\infty}p^{k}=p^{*}$, and we can prove the convergence of Algorithm \ref{Algorithm 1}.

\section{Proof of Theorem \ref{Theorem_convergence2}}
\label{AppendixB}
We assume the computation time of solving sub-problems for each BS is independent and follows exponential distribution, so we have the probability $p_s$ as follows
\begin{align}
p_s=Prob(s_k=s)={1\over S},
\end{align}
and we assume $p_{min}=min_{s}p_s>0$.

In order to prove the convergence of Algorithm \ref{Algorithm 2}, we first introduce the following Lemma \cite{peng2016arock}.
\begin{lemma}
We take the same settings of $\tau$ and J(k) in \cite{peng2016coordinate}, so for $\forall v^{*}\in \mathcal{M}_{DRS}\circ V^{*}=V^{*}$ for given $V^{k}=\{ v^{0},\ldots, v^{k} \}$ and $\sigma>0$, we have the following conditional expectation
\begin{eqnarray}\label{ine_consitional}
E(\|v^{k+1} - v^{*}\|^2|V^{k})&\leq& \|v^{k} - v^{*}\|^2+{\sigma\over S}\sum_{d\in J(k)}\|v^{d} - v^{d+1}\|^2\nonumber\\
&+&{1\over S}({{|J(k)|}\over \sigma}+{1\over {Sp_{min}}}-{1\over \eta_k})\|v^{k} - \bar{v}^{k+1}\|^2,
\end{eqnarray}
So we can observe that the upper bound of $E(\|v^{k+1} - v^{*}\|_2^2\|V^{k})$ only depends on $v^{*}$ and $V^{k}$.
\begin{IEEEproof}
\begin{eqnarray}
E(\|v^{k+1} - v^{*}\|^2|V^{k})&=&E(\|v^{k}-{\eta_k\over {Sp_{min}}}S_{s_k}\hat{v}^{k} - v^{*}\|^2|V^{k})\nonumber\\
&=&\|v^{k} - v^{*}\|^2+{2\eta_k\over S}\langle S\hat{v}^{k}, v^{*}-v^{k}\rangle+{\eta_k^{2}\over S^{2}}\sum _{s=1}^{S}{1\over p_s}\|S_s\hat{v}^{k}\|^2,
\end{eqnarray}
For $\sum _{s=1}^{S}{1\over p_s}\|S_s\hat{v}^{k}\|^2 \leq {k\over\eta_k^{2} p_{min}}\|v^{k} - \bar{v}^{k+1}\|^2$, and
\begin{eqnarray}
\langle S\hat{v}^{k}, v^{*}-v^{k}\rangle \leq -{1\over{2\eta_k^{2}}}\|v^{k} - \bar{v}^{k+1}\|^2+{{|J(k)|}\over {2\sigma\eta_k}}\|v^{k} - \bar{v}^{k+1}\|^2+{\sigma\over {2\eta_k}}\sum_{d\in J(k)}\|v^{d} - v^{d+1}\|^2.
\end{eqnarray}
This can prove inequality (\ref{ine_consitional}).
\end{IEEEproof}
\end{lemma}
\begin{lemma}
Let $\bv^{k}=(v^{k}, v^{k-1},\ldots, v^{k-\tau}),k\geq0$ and $\bv^{*}=(v^{*}, v^{*},\ldots, v^{*})\in \bV^{*}$, and $v^{k}=v^{0}$ for $k<0$, we have
\begin{eqnarray}
\xi_k(\bv^{*}):=\|\bv^{k} - \bv^{*}\|^2=\|v^{k} - v^{*}\|^2+\sqrt{p_{min}}\sum_{i=k-\tau}^{k-1}(i-(k-\tau)+1)\|v^{i} - v^{i+1}\|^2.
\end{eqnarray}
For $\forall \bv^{*}\in \bV^{*}$, we have
\begin{eqnarray}\label{ine_consitional2}
E(\xi_{k+1}(\bv^{*})|V^{k})\leq \xi_k(\bv^{*})+{1\over S}({2\tau\sqrt{k}\over {S\sqrt{p_{min}}}}-{1\over \eta_k})\|\bar{v}^{k+1}-v^{k}\|^2.
\end{eqnarray}
\begin{IEEEproof}
Let $\sigma=S\sqrt{p_{min}\over k}$, we have
\begin{eqnarray}
E(\xi_{k+1}(\bv^{*})|V^{k})&=&E(\|v^{k+1} - v^{*}\|^2|V^{k})+{\sigma\tau \over S}E({\eta_k^{2}\over {S^{2}p_{s_k}^{2}}}\|S_{s_k}\hat{v}^{k}\|^2|V^{k})\nonumber\\
&+&\sigma\sum _{i=k+1-\tau}^{k-1}{i-(k-\tau)\over S}\|v^{i} - v^{i+1}\|^2\nonumber\\
&\leq&\|v^{k} - v^{*}\|^2+{1\over S}({\tau\over \sigma}+{{\sigma\tau k}\over {S^{2}p_{min}}}+{k\over {Sp_{min}}}+{1\over\eta_k})\|v^{k} - \bar{v}^{k+1}\|^2\nonumber\\
&+&{\sigma\over S}\sum _{i=k-\tau}^{k-1}\|v^{i} - v^{i+1}\|^2+\sigma\sum _{i=k+1-\tau}^{k-1}{i-(k-\tau)\over S}\|v^{i} - v^{i+1}\|^2\nonumber\\
&=&\xi_k(\bv^{*})+{1\over S}({2\tau\sqrt{k}\over S\sqrt{p_{min}}}+{k\over {Sp_{min}}}+{1\over\eta_k})\|v^{k} - \bar{v}^{k+1}\|^2.
\end{eqnarray}
This can prove inequality (\ref{ine_consitional2}).
\end{IEEEproof}
\end{lemma}

Combing inequalities (\ref{ine_consitional}) and (\ref{ine_consitional2}), we can derive that $(\bV^{k})_{k\geq0}$ will converge to $\bV^{*}$ with probability of 1, and that finishes the proof of Throrem~\ref{Theorem_convergence2}.

\bibliography{reference}
\bibliographystyle{IEEEtran}
\end{document}